\documentclass[journal]{IEEEtran}
\ifCLASSINFOpdf
\else
\fi
\usepackage[font={footnotesize}]{caption}
\usepackage{subcaption}
\usepackage{balance}
\usepackage{algorithmic}
\usepackage{graphicx}
\usepackage{textcomp}
\usepackage{tikz}
\usepackage{pgfplots} 
\usepackage{tikzscale}
\usepackage{amsmath,amssymb,amsfonts}
\usepackage{accents} 
\usepackage{soul}
\usepackage{comment}
\usepackage[titlenumbered,ruled]{algorithm2e}
\usepackage{listings}
\usepackage{url}
\usepackage{xurl}
\usepackage{hyperref}
\usetikzlibrary{arrows.meta}
\usepackage{dashrule} 

\usepackage{bm}
\usepackage{cite}
\usepackage{multicol}
\usepackage{xcolor,colortbl}
\usepackage{diagbox}
\usepackage{multirow}
\usepackage{hanging}
\usepackage{ragged2e}
\usepackage{romannum}
\usepackage{xcolor}

\newcommand{\rom}[1]{\MakeUppercase{\romannum{#1}}}
\newcommand{\aline}

\begin{document}
%
\title{Gesture Recognition from body-Worn RFID under Missing Data}
\author{Sahar~Golipoor,
        Richard~T.~Brophy, Ying~Liu, Reza~Ghazalian, and Stephan~Sigg
\thanks{We acknowledge funding by the European Union in the frame of the Horizon Europe EIC project SUSTAIN (project no. 101071179), and Holden (project no. 101099491). Views and opinions expressed are those of the authors and do not necessarily reflect those of the European Union.}
\thanks{Sahar Golipoor, Ying Liu, Reza Ghazalian and Stephan Sigg are with the Department of Information and Communications Engineering, Aalto University, Espoo, 02150 Finland  (e-mail: \{sahar.golipoor, ying.2.liu, ext-reza.ghazalian, stephan.sigg\}@aalto.fi).}

\thanks{Richard T. Brophy is with the University of Pittsburgh, Pittsburgh, PA 15260, USA  (e-mail: brophy.ricky@gmail.com)}

}

\markboth{Journal of \LaTeX\ Class Files,~Vol.~14, No.~8, August~2015}%
{Shell \MakeLowercase{\textit{et al.}}: Bare Demo of IEEEtran.cls for IEEE Journals}
%



\maketitle
\sethlcolor{blue!50!white}
\begin{abstract}
We explore hand-gesture recognition through the use of passive body-worn reflective tags.
A data processing pipeline is proposed to address the issue of missing data.
Specifically, missing information is recovered through linear and exponential interpolation and extrapolation.
Furthermore, imputation and proximity-based inference are employed.
We represent tags as nodes in a temporal graph, with edges formed based on correlations between received signal strength (RSS) and phase values across successive timestamps, and we train a graph-based convolutional neural network that exploits graph-based self-attention.
The system outperforms state-of-the-art methods with an accuracy of 98.13\% for the recognition of 21~gestures. We achieve 89.28\% accuracy under leave-one-person-out cross-validation. 
We further investigate the contribution of various body locations on the recognition accuracy. Removing tags from the arms reduces accuracy by more than 10\%, while removing the wrist tag only reduces accuracy by around 2\%. Therefore, tag placements on the arms are more expressive for gesture recognition than on the wrist.
\end{abstract}


%
\IEEEpeerreviewmaketitle

\section{Introduction}
%
%
%
%
\IEEEPARstart{R}{adio} based human sensing analyzes electromagnetic signals reflected from people or objects to infer position, gestures, vital signs, and emotional states, using technologies such as Wi-Fi~\cite{miao2025wi}, radar~\cite{kong2024survey}, RFID~\cite{zahid2024comprehensive}, Bluetooth~\cite{baba2025human}, LoRa~\cite{obiri2024survey}, or cellular systems~\cite{yin2024systematic}.
Compared to vision-based methods, radio sensing has advantages, such as robustness to lighting conditions and occlusion.
Although radio sensing systems can raise distinct privacy concerns due to their ability to sense through obstacles and detect fine-grained movements, they do not capture identifiable visual features.
The applications of radio sensing extend to assisted living, healthcare and human-robot interaction~\cite{wen2024survey}.

Radio-based gesture recognition can be categorized into environmental and wearable methods.
Environmental approaches (e.g. Wi-Fi, radar, etc) often rely on changes in signal propagation caused by movements within the sensing area.
Reflections, distortions, or disruptions of an electromagnetic signal are interpreted to characterize gestures.
Environmental methods are well suited for applications targeting ease of use.
Wearable gesture recognition systems utilize body-worn sensors (e.g. backscatter tags)~\cite{azarfar2024real}.

Environmental backscatter systems deploy tag arrays within the environment, and individuals perform gestures within the detection range of these arrays
~\cite{zhang2024sign,xu2023rf,merenda2022edge}.
Wearable systems place the tags on the body~\cite{azarfar2024hand} to enable differentiation between body parts.
Wearable systems also allow individuals to evade sensing, e.g., for privacy reasons.

We explore the use of body-mounted Electronic Product Code (EPC) tags for gesture recognition and propose zero-padding and adaptive interpolation to address data loss (e.g., missing tags or samples) caused by polarization mismatch.
To further improve recognition accuracy, we propose algorithmic interpolation and imputation strategies.
Our contributions are
\begin{itemize}
\item a gesture recognition system from body-worn backscattering tags and its validation in three environments with data from~17 subjects.
\item linear and exponential step-wise interpolation techniques as well as an imputation algorithm based on intra-class proximity and proximity-based inference.
\item data representations to capture the correlation between received signal strength (RSS) and phase values associated with different EPCs across successive timestamps, enabling efficient propagation through graph-based self-attention convolution.
\end{itemize}

\section{Related Work}
Popular examples of environmental systems for motion recognition are Wi-Fi, radar and RFID.
Particularly, the channel state information (CSI) correlates with movement relative to a receiver.~\cite{regani2022gwrite}.
Combined with deep learning, high recognition accuracies are achieved~\cite{yao2023human,chen2024wignn}.
A limitation of CSI-based systems is their domain dependence.
To mitigate this, arranging antennas circularly around the sensing area may support orientation-agnostic recognition~\cite{qin2024direction}.
Furthermore, self-attention has been used to overcome domain dependence in CSI-based systems~\cite{gu2023attention,gu2022wigrunt}.

In contrast, radar-based systems exploit range-doppler information, as well as spatial diversity in an antenna array to obtain information on the distance, angle and velocity of objects relative to the radar~\cite{jin2023interference,yu2024mmwave,wu2024lightweight}.
Systems based on in-phase and quadrature components~\cite{jin2024rodar}, image processing over the time-frequency spectrum~\cite{qiao2024simple}, as well as point-cloud based systems~\cite{salami2022tesla} have been proposed.

RFID-based systems use machine learning to interpret fluctuations in RSS and signal phase, utilizing an array of tags.~\cite{zhang2024sign,xu2023rf,dian2020towards}.
Support vector machines, random forest, or CNN-based reasoning are prominent classification algorithms~\cite{merenda2022edge,zhang2022real,wang2018multi}, but also Dynamic Time Warping (DTW) to group temporal patterns with a template database~\cite{zou2016grfid}.

A common challenge for environmental systems is that signal reflections are superimposed from the complete environment and that relevant signal components first have to be isolated.
Furthremore, these systems raise privacy concerns since their operation is opaque and no easy means to avoid being captured by the system exist.


We propose to integrate reflective tags into clothing.
Therefore, relevant signals are easily separated from environmental reflections and avoidance of being captured is simply achieved by not wearing tags.

Body-worn sensors and deep learning have been widely used in health, human-computer interaction, and personalized medical care~\cite{nong2024intelligent,jeon2024applying,tang2024convolutional}.
Particularly, Inertial Measurement Units (IMUs) are common for motion recognition~\cite{saraf2023survey}.
Essential for a high recognition accuracy are the types of sensors and their placement on the human body~\cite{eddy2024discrete,mubibya2022improving,calatrava2023light}, which can be addressed, e.g. using transfer learning~\cite{li2023semg} or transformer-based structures~\cite{wang2024generalizations}.
Common challenges are the need for frequent charging or battery replacement~\cite{yang2024intelligent}.

We propose the use of passive reflective tags which overcomes this limitation.

Attaching RFID tags to objects has been employed for human-object interactions~\cite{li2015idsense,zhou2017design,shangguan2017enabling}.
By using orthogonally oriented antennas, object motion can be tracked~\cite{bu2018rf}.
Temporal gesture patterns can be distinguished using DTW~\cite{cheng2019air,xie2017multi} and rotation agnostic recognition is possible by normalizing the horizongtal rotation angle and radial distance~\cite{zhang2023rf}.
Coarse-grained gesture recognition utilizing machine learning and body-attached reflective tags has been presented in~\cite{yu2019rfid,golipoor2024environment}.

We propose integrating reflective tags into clothing, thereby allowing the system to distinguish between various body parts using RSS and phase signals.~\cite{golipoor2023accurate,golipoor2024rfid}.

Challenges of recognition systems using body-worn tags are frequent orientation changes of the tags, which may cause poor signal quality due to polarization mismatch~\cite{clarke2006radio}.
Furthermore, the anatomical structure and tissue composition of the human body affects signal reflection patterns~\cite{vasisht2018body}.

In this work, we employ a graph neural network to leverage the correlations between RSS and phase values from different tags across consecutive timestamps in a gesture pattern.
Furthermore, we address the problem of missing data.

\subsubsection*{Notation}
Vectors and matrices are denoted by lowercase and uppercase bold letters, respectively. The $i$-th element of vector $\mathbf{a}$ is represented by $[\mathbf{a}]_{i}$. For a matrix $\mathbf{A}$, $\mathbf{A}(:, i:j)$ selects all elements in all rows and columns $i$ through $j$ (inclusive), while $\mathbf{A}(i:j, :)$ selects all elements in rows $i$ through $j$ (inclusive) and all columns. The elements in the $\ell$-th row and $\ell$-th column are denoted by $\mathbf{A}(\ell, :)$ and $\mathbf{A}(:, \ell)$, respectively. For a vector $\mathbf{a}$, we use the notation $\mathbf{a}_{i:j}$ to denote the subsequence consisting of the elements from position $i$ to position $j$ (inclusive). The transpose operator is $(\cdot)^\top$. The $L_2$ norm (Euclidean norm for vectors, Frobenius norm for matrices) is denoted by $\| \cdot \|_2$. The cardinality of a set or the length of a vector is denoted by $|\cdot|$. The $L_0$ norm, $\|\mathbf{a}\|_0$, denotes the number of non-zero elements in $\mathbf{a}$. A column vector of length $Q$ with all elements equal to one is represented by $\mathbf{1}_Q$, and a column vector or matrix of size $Q$ with all elements equal to zero is represented by $\mathbf{0}_Q$. Assignment, where the value of $b$ replaces the value of $a$, is denoted by $a \leftarrow b$.
 In this paper, for two vectors $\mathbf{a}$ and $\mathbf{b}$, we define $\mathbf{a} \setminus \mathbf{b}$ as a vector containing the elements of $\mathbf{a}$ that are not present in $\mathbf{b}$, preserving their original order from $\mathbf{a}$. Note that we assume $\mathbf{b}$ is a sub-sequence of $\mathbf{a}$ for this operation. For example, if $\mathbf{a} = [1, 2, 3, 4]$ and $\mathbf{b} = [1, 4]$, then $\mathbf{a} \setminus \mathbf{b} = [2, 3]$. The function $\operatorname{argsort}(\mathbf{a})$ returns the indices that would sort the vector $\mathbf{a}$ in ascending order. The complement of a set $a$ is denoted by $a^c$, and $\emptyset$ denotes the empty set. For two sets $A$ and $B$, the notation $A \subseteq B$ means that $A$ is a subset of $B$.


\section{Overview of the measurement System}\label{overview}\label{SystemModel}
We utilize Alien AZ 9662 passive RFID tags as well as an Impinj Speedway R420 RFID reader and a circularly polarized Vulcan RFID PAR90209H antenna with a circular polarization gain of $9$~dBiC as well as elevation and azimuth beamwidth of $70^{\circ}$.
The reader is controlled by a laptop running the Impinj ItemTest software.
The RFID system operates in the 865MHz range at $30$ dBm.

The integrated circuit (IC) of the tag modulates its response to the continuous wave signal emitted by the reader by altering the load on its antenna to either backscatter or absorb the reader's signal, thereby producing reflection coefficient $\Gamma(t) \in \{0,1\}$.
The reflection coefficient is $\Gamma = \frac{Z_\text{Ant}-Z_\text{Load}(t)}{Z_\text{Ant}+Z_\text{Load}(t)}$, where $Z_\text{Ant}$ and $Z_\text{Load}(t)$ denote the antenna impedance (around $50\Omega$) and the load impedance controlled by the tag's IC.
$Z_\text{Load}= Z_\text{Ant}$ will result in $\Gamma=0$ (no signal reflected) while $Z_\text{Load}=0$ ($\Gamma=1$) causes the signal of the reader to be reflected.\footnote{In practice, due to imperfect load matching, the maximum value of the $\Gamma$ is close to $1$ and the minimum value is close to zero.}

We denote the basedband received signal at the reader by 
\begin{align}\label{eq:ym}
    y (t) &= \sum_{i=1}^{N_p} g_i G_a\sqrt{P_t}s(t) + \eta(t) ;\ \ \ \ y (t) \in \mathbb{C}.
\end{align}
In equation~(\ref{eq:ym}), $g_i\triangleq \frac {\lambda}{4\pi d_i}e^{\jmath \theta_i} $ denotes the channel gain of the $i$th round-trip path with $\theta_i\sim \mathcal{U}[0,2\pi)$, wavelength $\lambda$, path distance $d_i$, the number of multipath components $N_p$, antenna gain $G_a$, reader transmit power $P_t$ and the transmitted signal $s(t)$.
We model additive thermal noise at the reader as a zero-mean Gaussian $\eta(t)$ with variance $\sigma^2$.

The data received from the tag include EPCs, a timestamp, the signal phase, as well as the RSS.

\begin{figure*}
\begin{subfigure}{0.17\textwidth}
    \includegraphics[width=\linewidth]{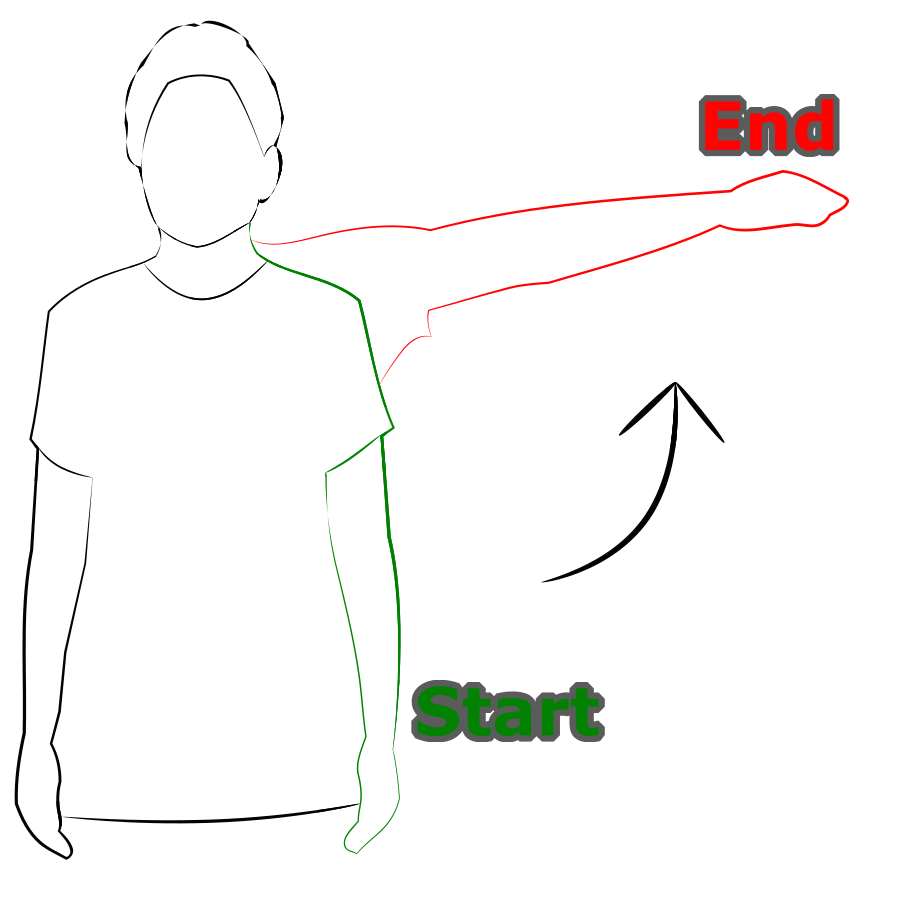}
    \caption{}
    \label{g}
\end{subfigure}
\begin{subfigure}{0.11\textwidth}
    \includegraphics[width=\linewidth]{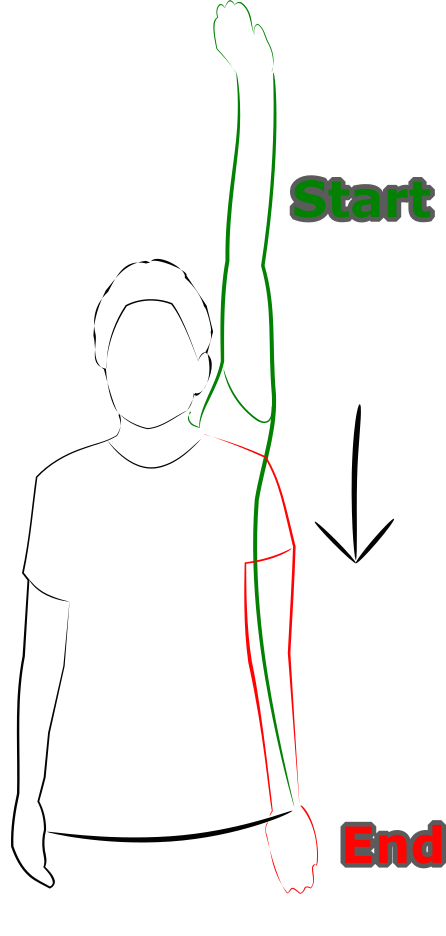}
    \caption{}
    \label{g}
\end{subfigure}
\begin{subfigure}{0.12\textwidth}
    \includegraphics[width=\linewidth]{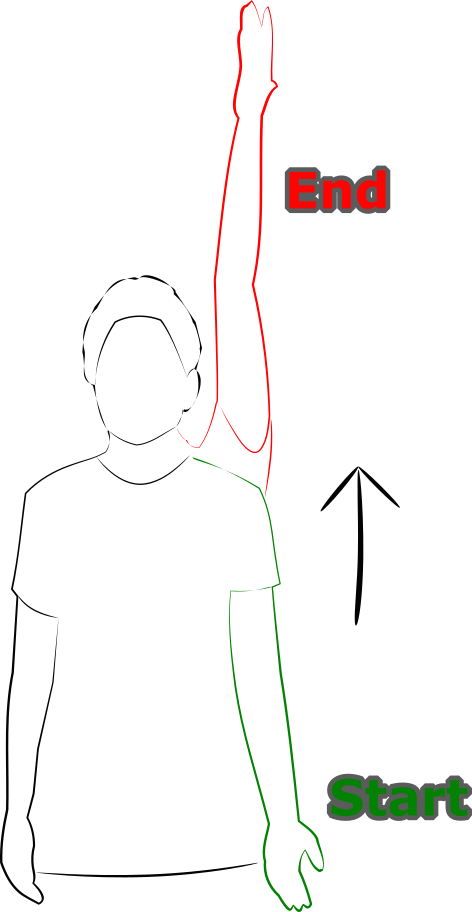}
    \caption{}
    \label{g}
\end{subfigure}
\begin{subfigure}{0.1\textwidth}
    \includegraphics[width=\linewidth]{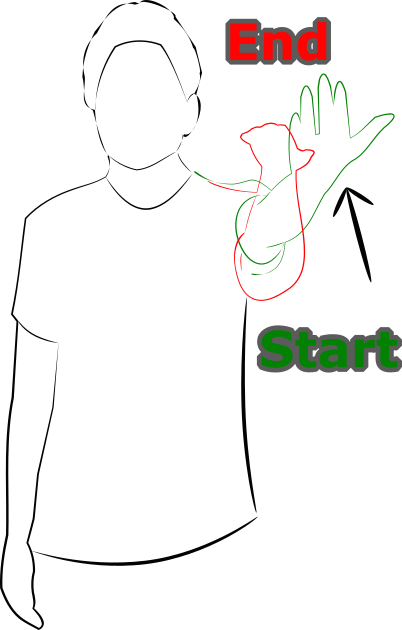}
    \caption{}
    \label{v}
\end{subfigure}
\begin{subfigure}{0.12\textwidth}
    \includegraphics[width=\linewidth]{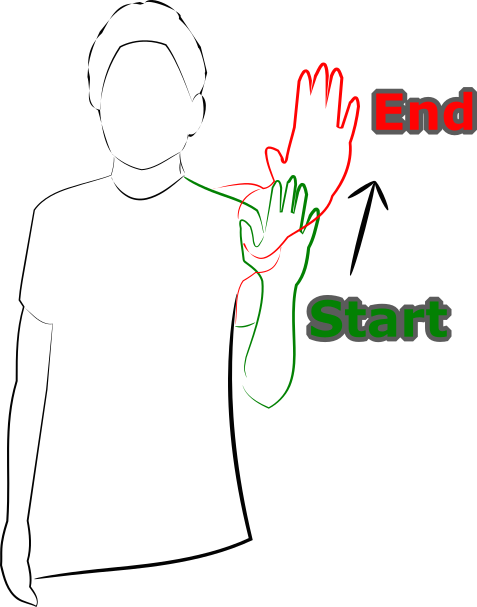}
    \caption{}
    \label{v}
\end{subfigure}
\begin{subfigure}{0.15\textwidth}
    \includegraphics[width=\linewidth]{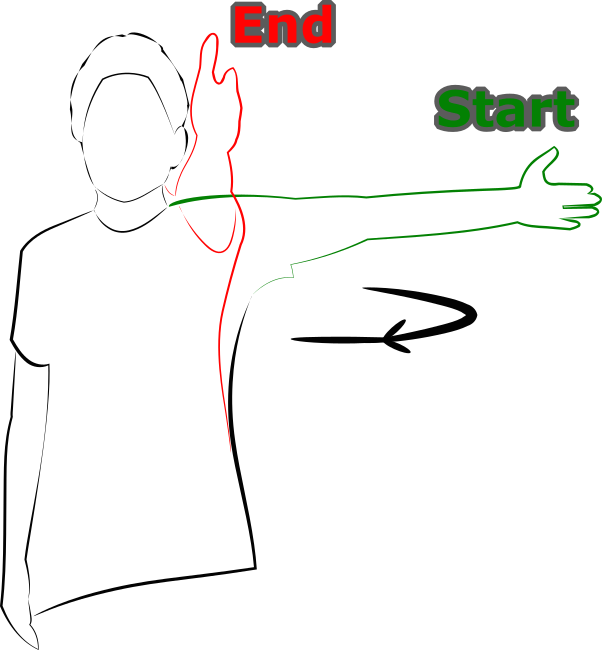}
    \caption{}
    \label{v}
\end{subfigure}
\begin{subfigure}{0.17\textwidth}
    \includegraphics[width=\linewidth]{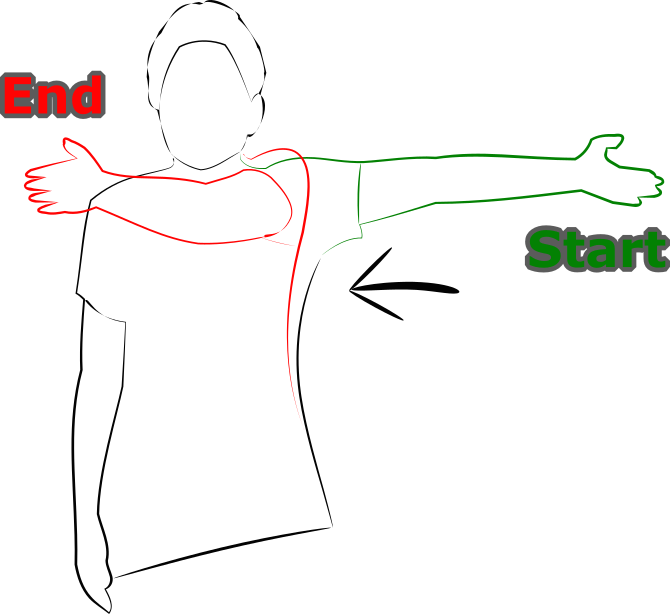}
    \caption{}
    \label{v}
\end{subfigure}
\begin{subfigure}{0.18\textwidth}
    \includegraphics[width=\linewidth]{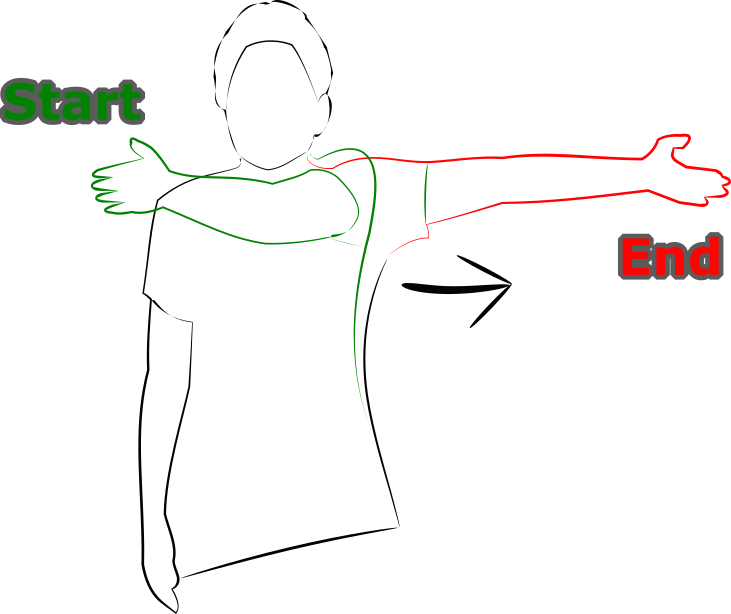}
    \caption{h}
    \label{v}
\end{subfigure}
\begin{subfigure}{0.11\textwidth}
    \includegraphics[width=\linewidth]{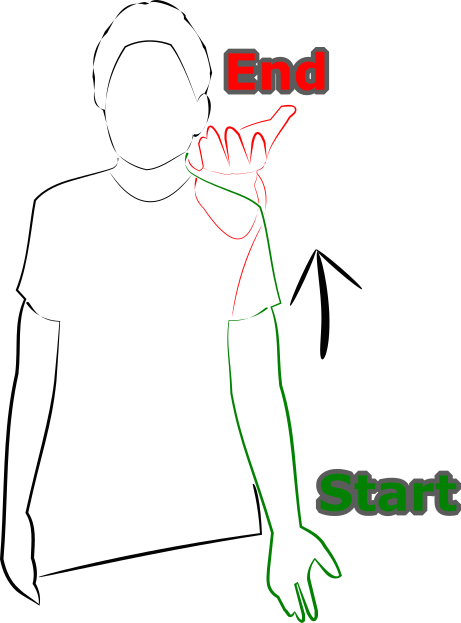}
    \caption{}
    \label{v}
\end{subfigure}
\begin{subfigure}{0.09\textwidth}
    \includegraphics[width=\linewidth]{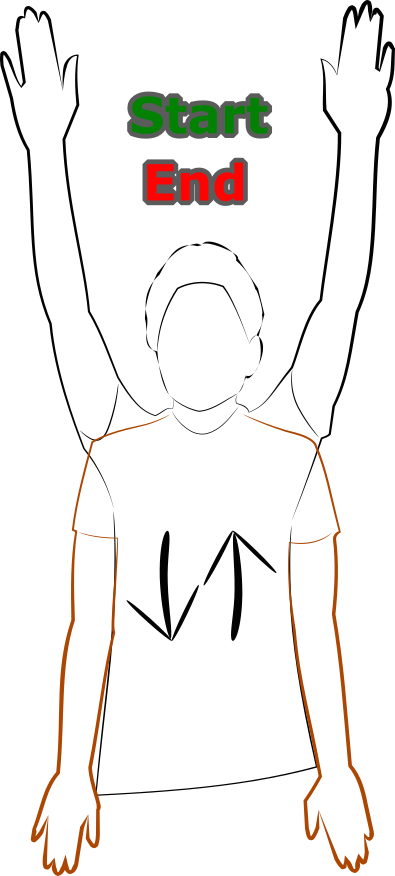}
    \caption{}
    \label{v}
\end{subfigure}
\begin{subfigure}{0.105\textwidth}
    \includegraphics[width=\linewidth]{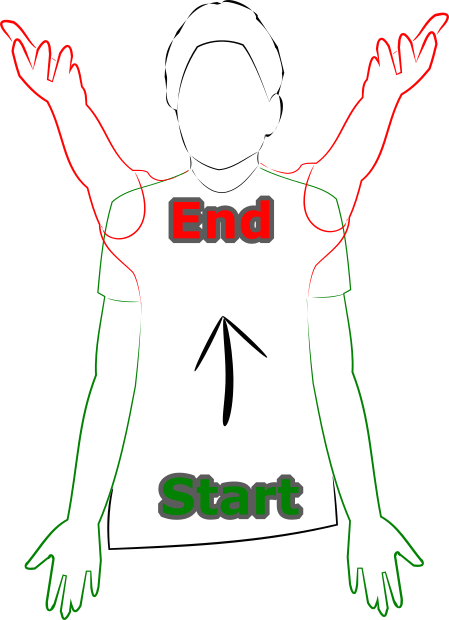}
    \caption{}
    \label{v}
\end{subfigure}
\begin{subfigure}{0.12\textwidth}
    \includegraphics[width=\linewidth]{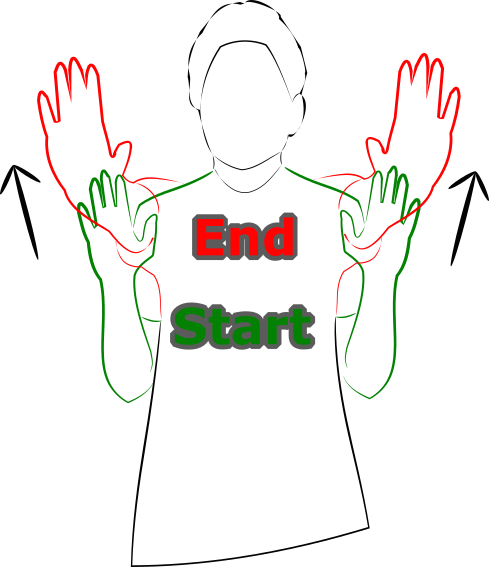}
    \caption{}
    \label{v}
\end{subfigure}
\begin{subfigure}{0.12\textwidth}
    \includegraphics[width=\linewidth]{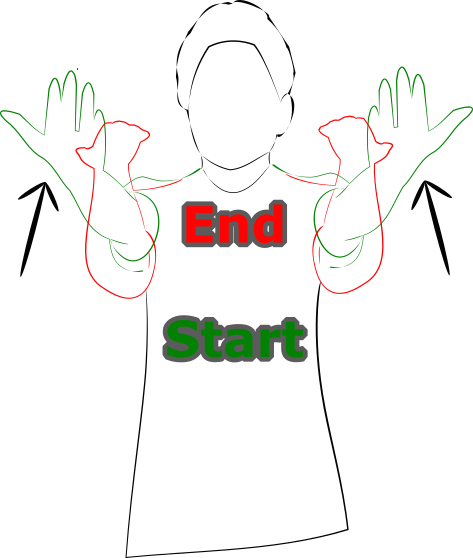}
    \caption{}
    \label{v}
\end{subfigure}
\begin{subfigure}{0.2\textwidth}
    \includegraphics[width=\linewidth]{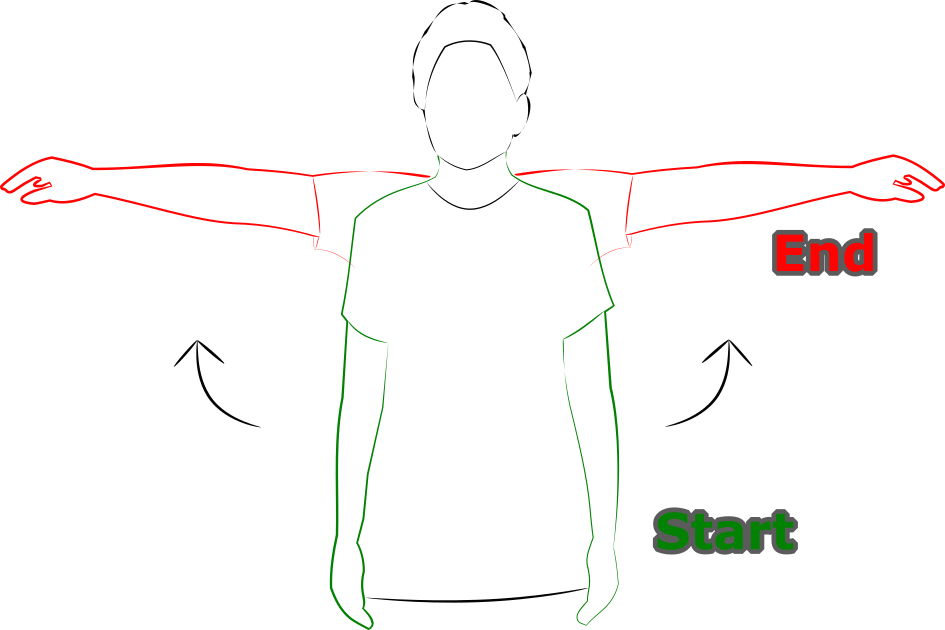}
    \caption{}
    \label{v}
\end{subfigure}
\begin{subfigure}{0.1\textwidth}
    \includegraphics[width=\linewidth]{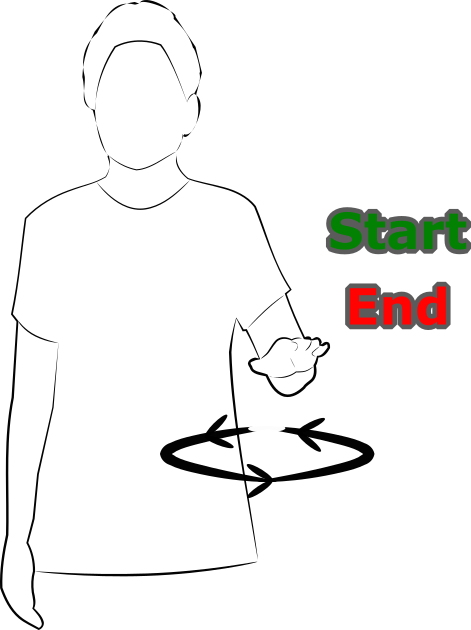}
    \caption{}
    \label{v}
\end{subfigure}
\begin{subfigure}{0.1\textwidth}
    \includegraphics[width=\linewidth]{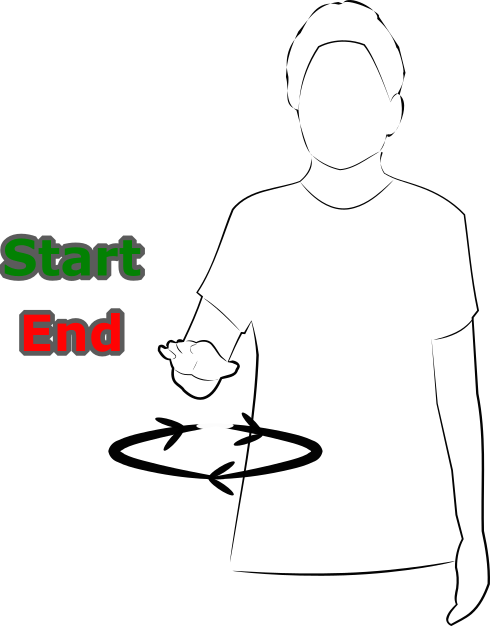}
    \caption{}
    \label{v}
\end{subfigure}
\begin{subfigure}{0.15\textwidth}
    \includegraphics[width=\linewidth]{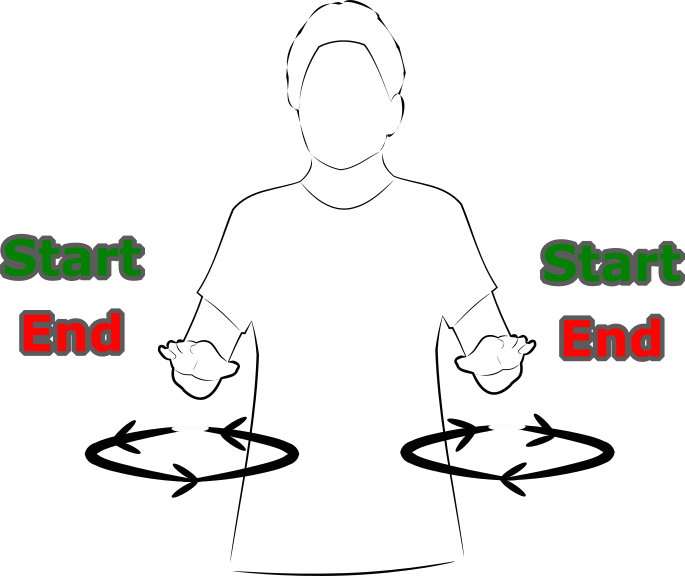}
    \caption{}
    \label{v}
\end{subfigure}
\begin{subfigure}{0.15\textwidth}
    \includegraphics[width=\linewidth]{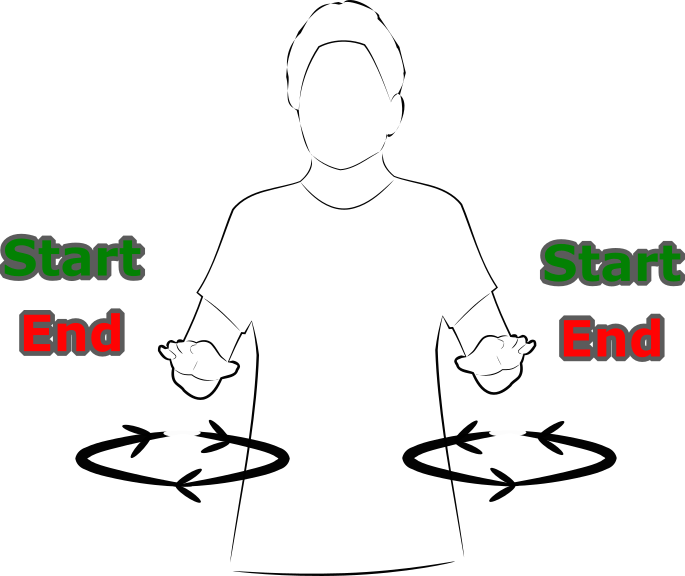}
    \caption{}
    \label{v}
\end{subfigure}
\begin{subfigure}{0.19\textwidth}
    \includegraphics[width=\linewidth]{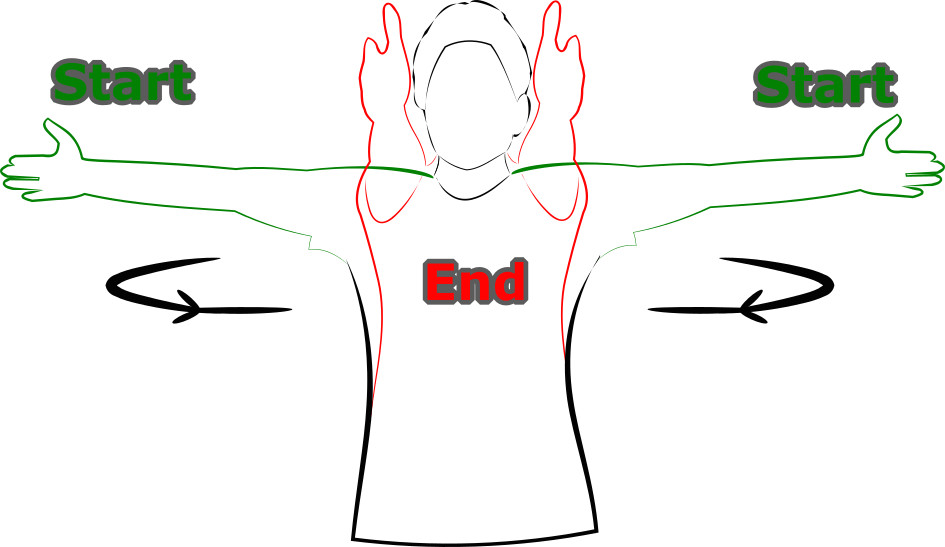}
    \caption{}
    \label{v}
\end{subfigure}
\begin{subfigure}{0.12\textwidth}
    \includegraphics[width=\linewidth]{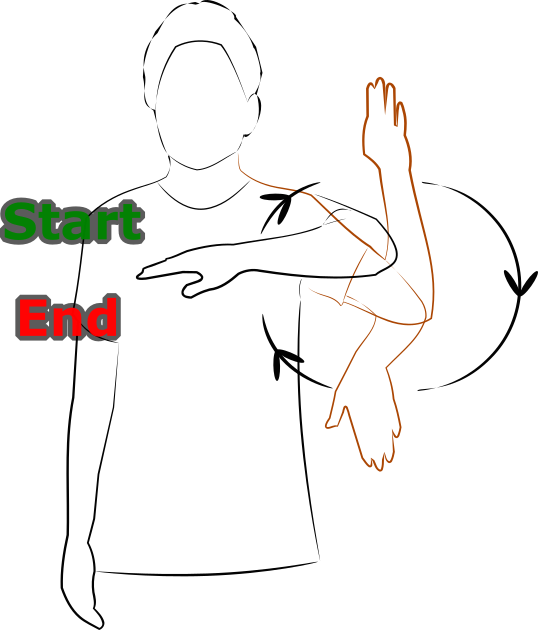}
    \caption{}
    \label{v}
\end{subfigure}
\begin{subfigure}{0.13\textwidth}
    \includegraphics[width=\linewidth]{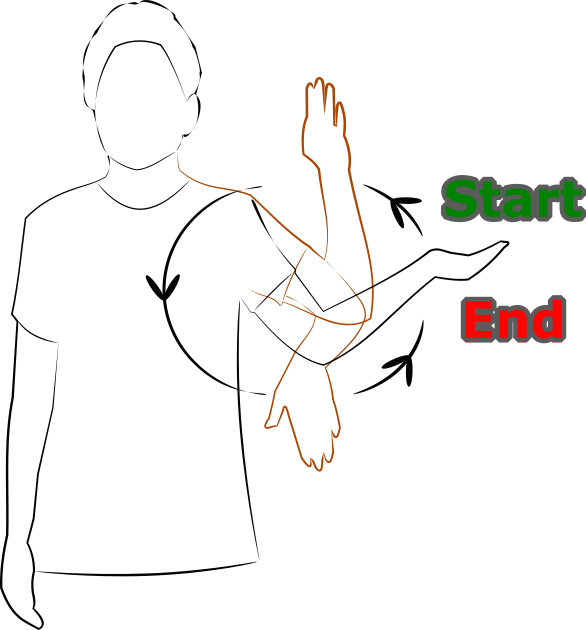}
    \caption{}
    \label{v}
\end{subfigure}
\caption{The gestures performed by participants are (a) lateral raise (1), (b) push down (2), (c) lift (3), (d) pull (4), (e) push (5), (f) lateral to front (6), (g) swipe right (7), (h) swipe left (8), (i) throw (9), (j) arm swings (10), (k) two-hand throw (11), (l) two-hand push (12), (m) two-hand pull (13), (n) two-hand lateral raise (14), (o) left-arm circle (15), (p) right-arm circle (16), (q) two-hand outward circles (17), (r) two-hand inward circles (18), (s) two-hand lateral to front (19), (t) clockwise circle (20), and (u) counterclockwise circle (21).}
\label{Gestures}
\end{figure*}
\begin{figure}
\centering
\includegraphics[width=0.8\linewidth]{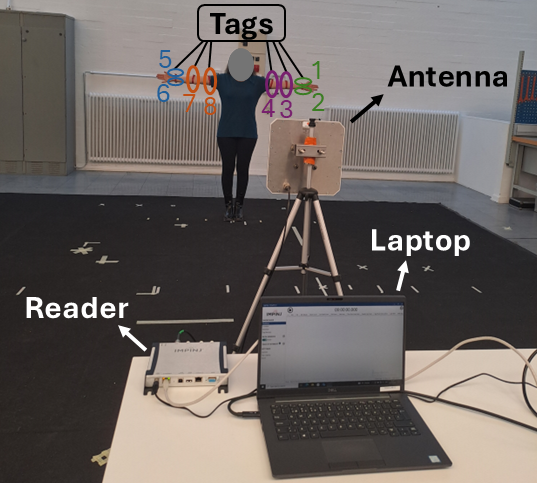}
\caption{Experimental setup. Eight passive RFID tags are attached to the subject, with phase and RSS data collected using an Impinj Speedway R420 reader and a circularly polarized antenna. Participants performed gestures at distances of 1.5m and 3m from the antenna.}
\label{SetUp}
\end{figure}

We collected 7,140 gesture samples from 17 participants (10 males, 7 females) with varying weight and height (155cm to 185cm), each performing 21 gestures 20 times (cf. Fig.~\ref{Gestures}).
Two tags were attached to the wrist of each hand, one below the elbow of each arm, and one on the upper arm of each hand as illustrated in Fig~\ref{SetUp}.
Gestures were performed at 3~and 1.5~meters distance from the antenna.
The duration of gestures varied due to varying gesture speed and gesture complexity.

\section{Data Preparation}\label{Data ProcessingMethod}
\begin{figure}
\centering
\includegraphics[width=0.9\linewidth]{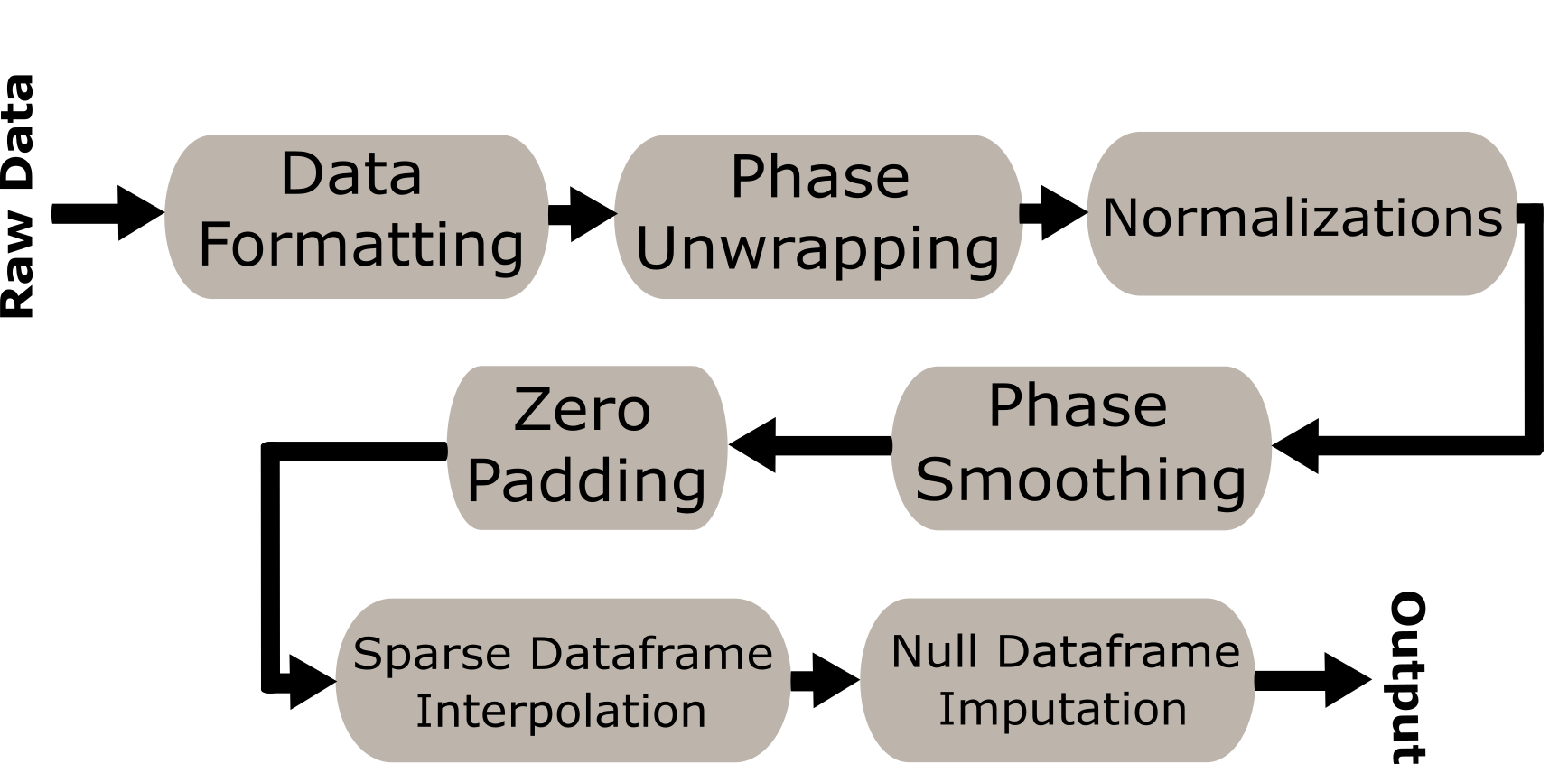}
\caption{Overview of the processing framework. Hardware-induced distortions are corrected, and missing data are addressed by stepwise interpolation and spatially guided data imputation.}
\label{Diagram}
\end{figure}

To facilitate further data analysis, the raw data from the reader is processed first.
Particularly, this regards, data formatting, phase unwrapping, normalizations, handling of missing samples, and interpolation.
The processing framework is illustrated in Fig~\ref{Diagram}.

Initially, we assign each EPC with a numerical identifier ranging from 1 to 8 (cf. Fig~\ref{SetUp}).
The RFID reader inventories tags sequentially, detecting only one EPC per timestamp. Additionally, due to movement, certain tags may become temporarily invisible during specific gestures caused by polarization mismatch between the antennas. As a result, EPCs have missing values at different timestamps. The data are initially ordered based on timestamps rather than EPCs. However, for subsequent processing tasks, it is necessary to organize the time-series data based on EPC.

Hence, the $i$-th gesture data $\mathbf{D}_i = [\mathbf{t}_i, \mathbf{e}_i, \mathbf{r}_i, \bm{\phi}_i] \in \mathbb{R}^{T \times 4}$,
 which corresponds to $i$-th gesture, consists of RSS and phase vectors $\mathbf{r}_i \in \mathbb{R}^T$ and $\bm{\phi}_i \in \mathbb{R}^T$ for the different EPCs $\mathbf{e}_i \in \mathbb{R}^T$ at various timestamps $\mathbf{t}_i = [t_1, t_2, \dots, t_T]^\top$, where $\mathbf{t}_i$ is the complete sequence of timestamps from the beginning to the end of the $i$th raw gesture, and $T$ is the total number of timestamp indices.

We sort $\mathbf{D}_i$ based on $\text{row}_\text{index} = \operatorname{argsort}(\mathbf{e}_i)$ to achieve $\tilde{\mathbf{D}}_i \triangleq  \mathbf{D}_i(\text{row}_\text{index}, :)= [\tilde{\mathbf{t}}_i, \tilde{\mathbf{e}}_i, \tilde{\mathbf{r}}_i, \tilde{\bm{\phi}}_i] \in \mathbb{R}^{T \times 4}$.
Particularly, $\tilde{\mathbf{D}}_i^n = [\tilde{\mathbf{t}}^n_i, \tilde{\mathbf{e}}^n_i, \tilde{\mathbf{r}}^n_i, \tilde{\bm{\phi}}^n_i] \in \mathbb{R}^{M_n \times 4}$ is the $n$th dataframe from the $i$th gesture data, where $\tilde{\mathbf{D}}_i \triangleq [(\tilde{\mathbf{D}}^1_i)^\top, \dots, (\tilde{\mathbf{D}}^N_i)^\top]^\top$, $M_n$ is the number of readings corresponding to the $n$th tag, and $\sum_{n=1}^N M_n = T$.
Therefore, a single gesture data point comprises eight dataframes, where each dataframe corresponds to an EPC and includes the associated RSS and phase values as well as a timestamp.

Next, phase unwrapping is applied followed by normalization.
We employ median absolute deviation (MAD) to normalize the signal phase due to its resilience to outliers:
\begin{eqnarray}
\beta_t &=& \text{median} \left( \left| [\tilde{\bm{\phi}}_i]_t - \text{median}(\tilde{\bm{\phi}}_i) \right| \right) \quad \forall t \in \{1\dots T\} \nonumber
\\
{[\tilde{\bm{\phi}}_i]_t} &=& \frac{[\tilde{\bm{\phi}}_i]_t-\text{median}(\tilde{\bm{\phi}}_i)}{\beta_t}.
 \label{eq: phaseNormalization}
\end{eqnarray}

The RSS data is normalized via Min-Max normalization:
\begin{equation}\label{eq: RSSNormalization}
[\tilde{\mathbf{r}}^n_i]_m \leftarrow \frac{[\tilde{\mathbf{r}}^n_i]_m - \min(\tilde{\mathbf{r}}^n_i)}{\max(\tilde{\mathbf{r}}^n_i) - \min(\tilde{\mathbf{r}}^n_i)} \quad \forall m= \{1, 2, \dots, M_n\}.
\end{equation}

Next, we employ Savitzky-Golay (S-G)~\cite{schafer2011savitzky}  and Gaussian filters~\cite{marr1980theory} to remove high-frequency noise:
\begin{equation}\label{eq: SGfilter}
    \tilde{\bm{\phi}}^n_i \gets \mathfrak{G}(\mathfrak{S}(\tilde{\bm{\phi}}^n_i)).
\end{equation}
In equation~(\ref{eq: SGfilter}), $\mathfrak{G} (.)$ and $\mathfrak{S}(.)$ represent the Gaussian and S-G filters.

For missing RSS and phase samples, we apply zero padding:
\begin{equation} \label{eq:phi_zero_padding}
\boldsymbol{\varphi}^n_i(t) = 
\begin{cases}
\tilde{\bm{\phi}}^n_i(t), & \text{if } t \in \tilde{\mathbf{t}}^n_i \\
0,                         & \text{if } t \in \tilde{\mathbf{t}}_i \setminus \tilde{\mathbf{t}}^n_i
\end{cases}
\end{equation}

\begin{equation} \label{eq:rss_zero_padding}
\boldsymbol{\rho}^n_i(t) = 
\begin{cases}
\tilde{\mathbf{r}}^n_i(t), & \text{if } t \in \tilde{\mathbf{t}}^n_i \\
0,                         & \text{if } t \in \tilde{\mathbf{t}}_i \setminus \tilde{\mathbf{t}}^n_i,
\end{cases}
\end{equation}
where \( \boldsymbol{\rho}^n_i \) and \( \boldsymbol{\varphi}^n_i \) represent the zero-padded RSS and phase data of the $n$-th dataframe of the $i$-th gesture.
This results in dataframes that either contain only zero values, \textit{null dataframe}, or contain a mixture of zero and non-zero values, \textit{sparse dataframe}. While the \textit{null dataframe} is addressed using within-class imputation methods in combination with spatial proximity imputation, the \textit{sparse dataframe} is processed using linear and exponential interpolation techniques.

\subsection{Interpolation}
In each \textit{sparse dataframe}, zero values occurring between two valid observations of phase and RSS are addressed through linear and exponential interpolation, respectively. However, in many cases, a dataframe may begin or end with a sequence of zeros, making interpolation infeasible due to the lack of two reference points. For dataframes with leading or trailing zero phase values, we apply linear extrapolation according to the first (last) two non-zero phase values, i.e.,
\begin{equation}\label{eq:lininterpol}
    \boldsymbol{\varphi}^n_i \leftarrow \mathcal{F}(\boldsymbol{\varphi}^n_i)
\end{equation}
where \( \mathcal{F}(.) \) performs the linear interpolation and extrapolation. This approach is similarly applied to RSS values, using exponential extrapolation to fill leading and trailing zero values, i.e.,
\begin{equation}\label{eq:exinterpol}
    \boldsymbol{\rho}^n_i \leftarrow \mathcal{G}(\boldsymbol{\rho}^n_i)
\end{equation}
where \( \mathcal{G}(.) \) performs the exponential interpolation and extrapolation. Finally, dataframes are resampled to $l_\text{rs}$.

\subsection{Data Imputation}
We applied within-class imputation along with spatial proximity imputation to fill the \textit{null dataframes}. For each reference gesture data, we identify the \textit{null dataframes} (i.e., those EPCs that are entirely missing during the inventory round) and the \textit{sparse dataframe} (i.e., dataframes with available EPCs that have already undergone interpolation in the previous step).

To address data imputation, we introduce two metrics. The first metric quantifies the similarity between two gesture samples using the Mean Euclidean Distance (MED).
Let $\bar{\mathbf{D}}^a_i\in \mathbb{R}^{k^a_i \times 4}$ represent a matrix that contains all sparse dataframes for the $i$-th gesture data, where the indices of their EPCs are contained in the set $\Delta_i$, with $|\Delta_i|=k^a_i$.
Similarly, for the $m$-th gesture data ($m \in {1, 2, \dots, M}$, $m \neq i$), with EPC indices corresponding to all sparse dataframes in the set $\Delta_m$, where $|\Delta_m| = k^a_m$ and $\Delta_i \subseteq \Delta_m$, we construct a matrix $\bar{\mathbf{D}}^{a^{\prime}}_m \in \mathbb{R}^{k^a_i \times 4}$ by vertically concatenating all dataframes with EPCs whose indices belong to the set $\Delta_i$.
The MED between gesture data $i$ and $m$ is then calculated as:
\begin{align}\label{eq:meanEc}
\kappa_{i,m} &= \frac{1}{k^a_i} \sum_{q=1}^{k^a_i} \| \bar{\mathbf{D}}^a_i(q,3:4) - \bar{\mathbf{D}}^{a^{\prime}}_m(q,3:4) \|_2.
\end{align}

The second metric considers the spatial proximity of the tags to perform data imputation. To do so, we define proximity sensors based on the placement of the tags on human clothing, as illustrated in Fig.~\ref{SetUp}. As such, the proximity sensors matrix can be defined $\mathbf{A}= [\mathbf{p}_1, \mathbf{p}_2, \mathbf{p}_3, \mathbf{p}_4]$ where $\mathbf{p}_1 =[1,2]^\top$, $\mathbf{p}_2 =[3,4]^\top$, $\mathbf{p}_3 =[5,6]^\top$, and $\mathbf{p}_4 =[7,8]^\top$. 
Now we consider cases where we need to perform imputation using the aforementioned metrics in the sequel.

\textbf{In the first case, the $i$-th reference gesture data contains only one null dataframe for the EPC with index $p$.} 
We first calculate all MED values between the $i$-th reference gesture data and the other gesture data within the same class, i.e., $\kappa_{i,m} \quad \forall m \neq i$ given in~\eqref{eq:meanEc}. We then define the matrix $\boldsymbol{K}$, where each row contains $\kappa_{i,m}$ in the first column and the corresponding index $m$ in the second column.
We identify the gesture data indices corresponding to the $\nu$ smallest values in the first column of $\boldsymbol{K}$ (i.e., $\boldsymbol{K}(:,1)$) by sorting it in ascending order. Let $v_1 \le v_2 \le \dots \le v_\nu$ be these $\nu$ smallest MED values, and let $\boldsymbol{\zeta} = [ \zeta_1, \zeta_2, \dots, \zeta_\nu ]^\top$ be the vector of the corresponding gesture data indices, such that, for example, $\kappa_{i,\zeta_1} = v_1$.

Then, we define the set $\Omega \triangleq \{\forall\zeta_j \mid \tilde{\mathbf{D}}^p_{\zeta_j}(:,3:4) \neq \mathbf{0}, j = 1, 2, \dots, \nu \}$, where $\tilde{\mathbf{D}}^p_{\zeta_j} (:,3:4) \neq \mathbf{0}$ indicates that the $\zeta_j$-th gesture in $\Omega$ contains a sparse dataframe for the EPC with index $p$. Finally, the missing value $\tilde{\mathbf{D}}^p_i(:,3:4)$ is imputed by averaging the RSS and phase values from the set $\Omega$:
\begin{equation}\label{eq:average miss}
\tilde{\mathbf{D}}^p_i(:,3:4) \leftarrow \frac{1}{|\Omega|} \sum_{\forall\zeta_j\in \Omega} \tilde{\mathbf{D}}^p_{\zeta_j}(:,3:4)
\end{equation}
If the MED metric cannot be computed because $\Omega = \emptyset$, the missing RSS and phase values are imputed using data from the nearest EPC within the same reference gesture, based on the proximity matrix $\mathbf{A}$.

\textbf{In the second case, the reference gesture data contains two null dataframes}.
To this end, we first determine whether the two null dataframes belong to the same EPC pair according to $\mathbf{A}$ (i.e., whether they are spatially close based on tag placement). If the null dataframes are not from the same pair, their corresponding dataframes are then filled using the approach mentioned in the first case (see~\eqref{eq:average miss}), if applicable. If the MED metric cannot be applied (due to a lack of available data), the null dataframes are imputed using the dataframes from the closest available EPC within the same reference gesture data based on $\mathbf{A}$.

In contrast, when the two null dataframes belong to the same EPC pair, a scenario with low probability as indicated in Fig.~\ref{fig:MisProbablity}, we initially assess the applicability of the MED metric for both. If the MED metric is applicable to both null dataframes, they are imputed using the methodology detailed in the first case (see~\eqref{eq:average miss}). If the MED metric is applicable to only one null dataframe (e.g., the first), it is imputed following the procedure described in the first case (see~\eqref{eq:average miss}). This imputed dataframe is then used to impute the second null dataframe, leveraging the spatial closeness of their EPCs. Finally, in the event that the MED metric is not applicable to either of null dataframes, the imputation is performed by averaging the values corresponding to the missing EPCs across all available gesture data within the same class. 
The second case can be generalized to instances that there are more than two null dataframes, on the condition that no proximity-paired EPCs are missing. Data imputation is summarized in \textit{Algorithm 2}, and our model is summarized in \textit{Algorithm 1}.

\begin{algorithm}[!t]
\caption{the Proposed Model}
\begin{algorithmic}[1]
\REQUIRE $i$-th raw gesture data $\mathbf{D}_i \in \mathbb{R}^{T \times 4}$, $N$, and $l_\text{rs}$
\textbf{Extract $i$-th sorted gesture data, $\tilde{\mathbf{D}}_i$, and dataframes $\tilde{\mathbf{D}}^n_i$ for all $n \in \{1, 2, \dots, N\}$}
\STATE  Obtain the row indices that sort the EPCs $\mathbf{D}_i(:, 2) = \mathbf{e}_i$ in ascending order: $\text{row}_\text{index} = \operatorname{argsort}(\mathbf{e}_i)$.
\STATE  Sort the rows of $\mathbf{D}_i$ based on $\text{row}_\text{index}$, i.e., $\tilde{\mathbf{D}}_i = \mathbf{D}_i(\text{row}_\text{index}, :)$.
\STATE  Define dataframes corresponding to each EPC number as: $\tilde{\mathbf{D}}_i \triangleq [(\tilde{\mathbf{D}}^1_i)^\top, \dots, (\tilde{\mathbf{D}}^N_i)^\top]^\top$.

\textbf{Phase Processing of gesture data and dataframes}
\FOR{$n = 1$ to $N$}
\STATE Perform phase unwrapping: $\tilde{\bm{\phi}}^n_i \gets \text{unwrap}(\tilde{\bm{\phi}}^n_i)$, and update $\tilde{\mathbf{D}}^n_i(:, 4) \leftarrow \tilde{\bm{\phi}}^n_i$.
\ENDFOR
\STATE  Update $\tilde{\mathbf{D}}_i$ based on the updated dataframes.
\STATE  Perform phase normalization over $\tilde{\bm{\phi}}_i$ based on Equation~\eqref{eq: phaseNormalization} and update $\tilde{\mathbf{D}}_i$ accordingly.
\FOR{$n = 1$ to $N$}
\STATE  Perform phase smoothing based on Equation~\eqref{eq: SGfilter} and update $\tilde{\mathbf{D}}^n_i(:, 4) \leftarrow \tilde{\bm{\phi}}^n_i$, and $\tilde{\mathbf{D}}_i$.
\ENDFOR

\textbf{RSS Normalization}
\FOR{$n = 1$ to $N$}
\STATE  Perform RSS normalization based on Equation~\eqref{eq: RSSNormalization}, update $\tilde{\mathbf{D}}^n_i(:, 3) \leftarrow \tilde{\mathbf{r}}^n_i$, and update $\tilde{\mathbf{D}}_i$. 
\ENDFOR

\textbf{Zero Padding $\tilde{\mathbf{D}}^n_i$ for $n \in \{1, 2, \dots, N\}$}
\FOR{$n = 1$ to $N$}
\STATE  Perform zero-padding using Equations~\eqref{eq:phi_zero_padding} and~\eqref{eq:rss_zero_padding}, resulting in $\tilde{\mathbf{D}}^n_i \leftarrow [\tilde{\mathbf{t}}_i, n\mathbf{1}_T, \boldsymbol{\rho}^n_i, \boldsymbol{\varphi}^n_i] \in \mathbb{R}^{T \times 4}$.
\STATE Perform zero padding validation:
\hspace{1em}\IF{$\|\boldsymbol{\rho}^n_i\|_0 < 2$ \OR $\|\boldsymbol{\varphi}^n_i\|_0 < 2$}
\STATE \hspace{1em} $\boldsymbol{\rho}^n_i \gets \mathbf{0}_T$ and $\boldsymbol{\varphi}^n_i \gets \mathbf{0}_T$
\ENDIF
\ENDFOR

\textbf{Perform interpolation/extrapolation and data imputation on $\tilde{\mathbf{D}}^n_i$ for all $n \in \{1, 2, \dots, N\}$}
\FOR{$n = 1$ to $N$}
\STATE Perform interpolation/extrapolation on $\boldsymbol{\rho}^n_i$ and $\boldsymbol{\varphi}^n_i$.
\IF{$\|\boldsymbol{\rho}^n_i\|_0 > 0$ \OR $\|\boldsymbol{\varphi}^n_i\|_0 > 0$}
\STATE \hspace{2em} Perform Equations~\eqref{eq:lininterpol} and~\eqref{eq:exinterpol}, and update $\tilde{\mathbf{D}}^n_i$ and $\tilde{\mathbf{D}}_i$.
\STATE  $\boldsymbol{\varphi}^n_i \leftarrow \text{Res}(\boldsymbol{\varphi}^n_i, l_\text{rs})$ and $\boldsymbol{\rho}^n_i \leftarrow \text{Res}(\boldsymbol{\rho}^n_i, l_\text{rs})$, $\tilde{\mathbf{D}}^n_i(:, 3:4) \gets [\boldsymbol{\rho}^n_i, \boldsymbol{\varphi}^n_i]$, and update $\tilde{\mathbf{D}}_i$.
\ELSE
\STATE \hspace{2em} Skip interpolation/extrapolation and proceed to data imputation (Algorithm 2).
\ENDIF
\ENDFOR

\textbf{Return:} Refined processed $i$-th gesture data $\tilde{\mathbf{D}}_i$.
\end{algorithmic}
\end{algorithm}
\begin{algorithm}[!t]
\caption{\textit{Null Dataframe} Imputation}
\begin{algorithmic}[1]
\REQUIRE All null dataframes $\tilde{\mathbf{D}}^n_i$ where $n \in \Delta^c_i$ (set of missed EPC numbers, $|\Delta^c_i| = \epsilon_i$), all gesture data sets in the training phase, available EPC numbers in the $i$-th gesture data: $\Delta_i$, Proximity sensor matrix $\mathbf{A}$, and the number of nearest gesture data to consider: $\nu$.
\STATE Extract the matrix of sparse dataframes for the $i$-th and $m$-th gesture data: $\bar{\mathbf{D}}^a_i, \bar{\mathbf{D}}^{a^{\prime}}_m \in \mathbb{R}^{k^a_i \times 4}$, where $\Delta_i \subseteq \Delta_m$.
\STATE Calculate $\kappa_{i,m}$ for all $i, m \in \{1, 2, \dots, M\}$, and $m \neq i$, based on Equation~\eqref{eq:meanEc}. Store $\kappa_{i,m}$ and $m$ in the first and the second column of matrix $\boldsymbol{K}$, respectively.
\STATE Find the $\nu$ row indices corresponding to the $\nu$ smallest values, $\mathcal{I}_\nu \triangleq [\operatorname{argsort}(\boldsymbol{K}(:, 1))]_{1:\nu}$, and store the indices in: 
$\boldsymbol{\zeta} \triangleq\boldsymbol{K}(\mathcal{I}_\nu, 2)$
\STATE Define the set of nearest neighbor gesture data indices $\Omega \triangleq \{ \forall\zeta_j \mid j = 1, 2, \dots, \nu \}$, where $\zeta_j \triangleq [\boldsymbol{\zeta}]_j$.
\FOR{$\forall \mu \in \Delta^c_i$}
\STATE Initialize the set of candidate gesture data for imputation: $\Omega_{\mu} \leftarrow \Omega$.
\STATE Update $\Omega_{\mu} \gets \{ \forall\zeta_j \in \Omega \mid \tilde{\mathbf{D}}^{\mu}_{\zeta_j}(:,3:4) \neq \mathbf{0} \}$.
\IF{$\Omega_{\mu} \neq \emptyset$}
\STATE Impute the null dataframe for the $\mu$-th EPC in the $i$-th gesture data $\tilde{\mathbf{D}}^{\mu}_i(:, 3:4)$ by averaging over the data in $\Omega_{\mu}$ according to Equation~\eqref{eq:average miss}, and update $\tilde{\mathbf{D}}_i$.
\ELSE
\IF{$\tilde{\mathbf{D}}^{\delta}_i \neq \mathbf{0}$, where $\delta$ and $\mu$ belong to the same column of matrix $\mathbf{A}$}
\STATE Impute: $\tilde{\mathbf{D}}^{\mu}_i \gets \tilde{\mathbf{D}}^{\delta}_i$, and update $\tilde{\mathbf{D}}_i$.
\ELSE
\STATE Perform Equation~\eqref{eq:average miss} over all gesture data that have available $\mu$-th EPC, and update $\tilde{\mathbf{D}}_i$.
\ENDIF
\ENDIF
\ENDFOR

\textbf{Return:} Refined version of the $i$-th dataframe $\tilde{\mathbf{D}}_i$.
\end{algorithmic}
\vspace{0.5em}
\textbf{Note 1:} Algorithm 2 is applicable when the number of missing EPCs $\epsilon_i \leq 4$ and these missing EPCs are not proximity tags.

\textbf{Note 2:} Line 14 is applicable only during training, where this scenario's probability is low (see Fig.~\ref{fig:MisProbablity}). During testing, only Line 12 is used, which is practical (see Fig.~\ref{fig:MisProbablity}).

\end{algorithm}

\section{Attention-Based Graph Learning}\label{Learning}
\begin{figure}
\centering
\includegraphics[width=0.9\linewidth]{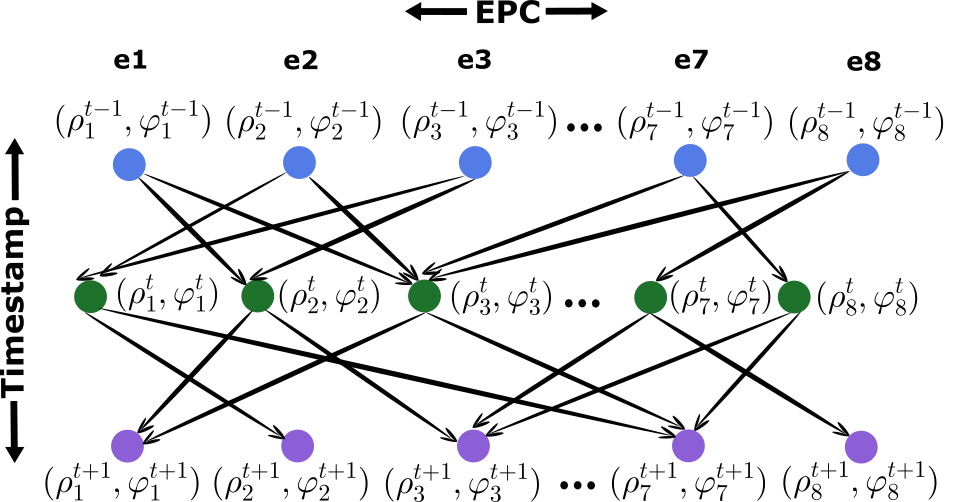}
\caption{Graph generation based on $(\rho_{i}, \varphi_{i})$ values across consecutive timestamps. For each EPC at a given timestamp, directed edges are formed by connecting to the nearest neighbors in the preceding timestamp using a temporal graph K-NN approach.}
\label{Graph}
\end{figure}

For the classification, we employ the graph-based convolutional neural network introduced in~\cite{salami2022tesla}. We use temporal K-nearest neighbors (K-NN)~\footnote{We use Euclidean distance as the metric to determine neighbors in the K-NN.} to form a graph where each EPC (tag) is represented as a node, and edges are established between nodes based on the proximity of their RSS and phase values across consecutive timestamps\footnote{In this study, with a large temporal hyperparameter, K-NN is applied to find neighbors between consecutive timestamps. Specifically, for two EPCs $i$ and $j$, an edge is formed if the signal feature of EPC $i$ at time $t-1$, $(\rho_{i}^{t-1}, \varphi_{i}^{t-1})$, is among the $K$ nearest neighbors of the signal feature of EPC $j$ at time $t$, $(\rho_{j}^{t}, \varphi_{j}^{t})$, in the feature space.}.
As illustrated in Fig.~\ref{Graph}, the framework connects each EPC to its most similar counterparts at the next timestamp, thereby capturing meaningful temporal correlations and enabling efficient feature propagation across the graph.

Afterward, it employs graph processing including message propagation, self-attention, and an aggregation function. In the message propagation step, node $i$ receives information from node $j$, where the information is a function of the features of these nodes. This function is implemented by a multilayer perceptron. To focus on the most relevant neighbors, a self-attention mechanism is incorporated. This mechanism computes attention weights for each neighbor $j$ of node $i$, scaling the influence of the corresponding message. The updated feature vector for node $i$ at the next layer is then obtained by aggregating the weighted messages and the node's previous feature\footnote{The aggregation function can be mean or summation.}.


To facilitate model training, the input data is structured into a four-dimensional tensor $\boldsymbol{\mathcal{T}} \in \mathbb{R}^{B \times T \times N \times D}$ where $B$ denotes the total number of gesture samples, $T$ is the number of timestamps per EPC in each gesture sample, $N$ represents the number of EPCs, and $D$ is the feature dimension of each EPC (i.e., RSS and phase).
In this work, the tensor shape used is ($T=30$, $N=8$, $D=2$), corresponding to 30 timestamps, 8 EPCs, and two-dimensional feature vectors per node.

\section{Evaluation}
\begin{figure*}
\centering
\begin{subfigure}{0.24\textwidth}
    \centering
    \includegraphics[width=\linewidth]{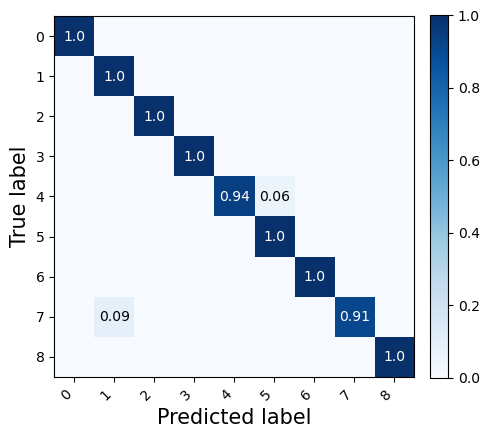}
    \caption{Within-subject; Accuracy=98.27\%}
    \label{EasyWithin}
\end{subfigure}
 \hfill
\begin{subfigure}{0.24\textwidth}
    \centering
    \includegraphics[width=\linewidth]{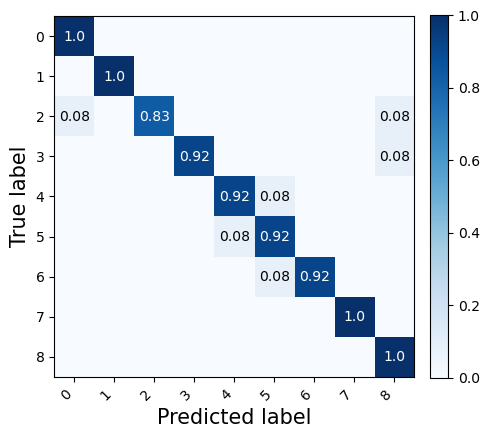}
    \caption{Leave-one-person-out; Acc=94.44\%}
    \label{EasyLopo}
\end{subfigure}
\hfill
\begin{subfigure}{0.24\textwidth}
    \centering
    \includegraphics[width=\linewidth]{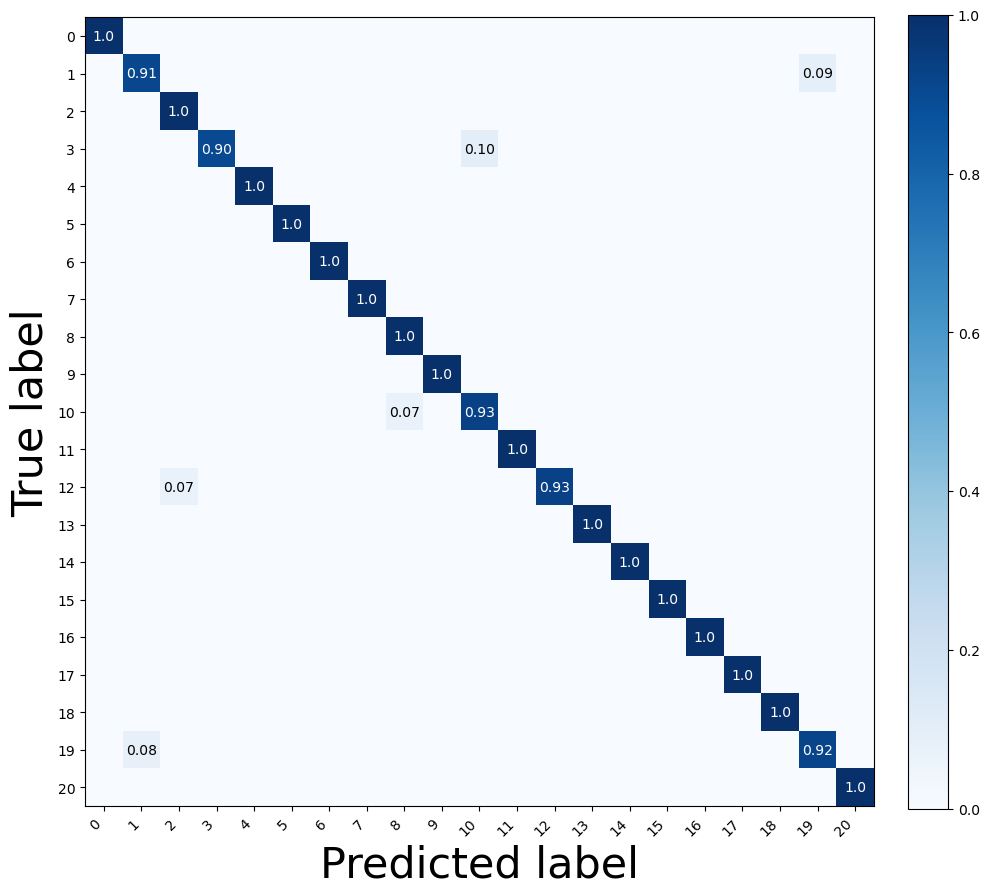}
    \caption{Within-subject; Accuracy=98.13\%}
    \label{ComplexWithin}
\end{subfigure}
 \hfill
\begin{subfigure}{0.24\textwidth}
    \centering
    \includegraphics[width=\linewidth]{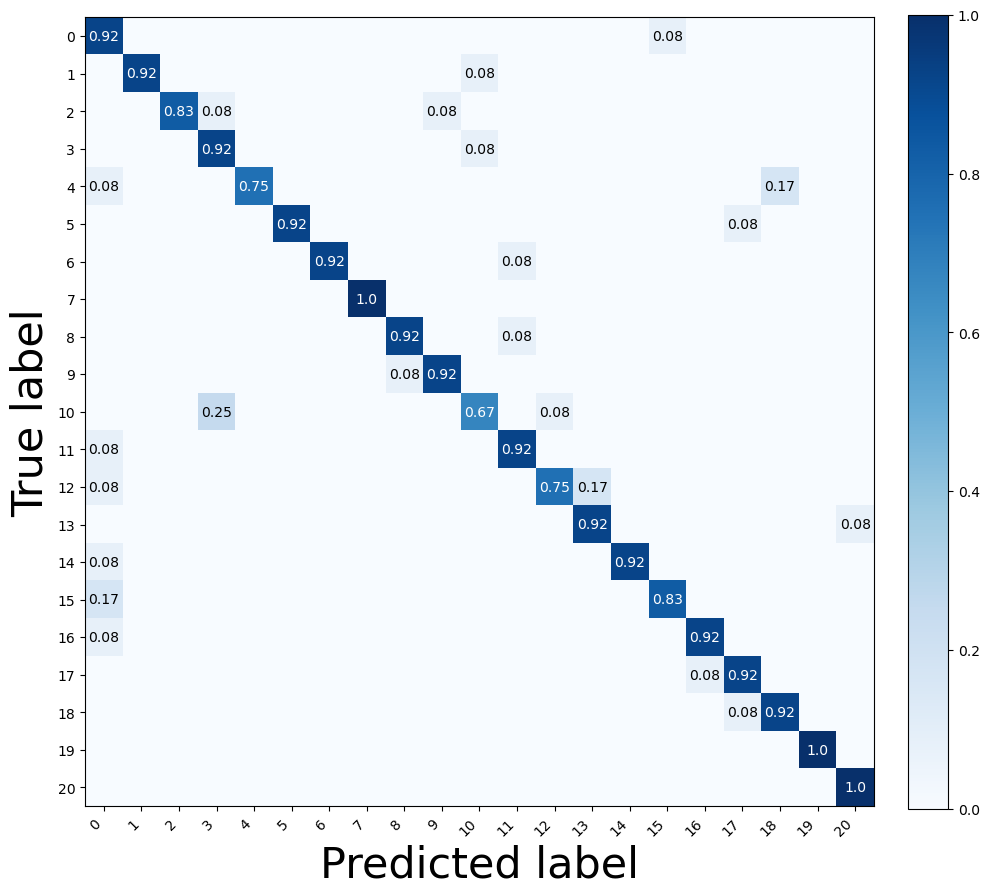}
    \caption{Leave-one-person-out; Acc=89.28\%}
    \label{ComplexLopo}
\end{subfigure}
\caption{Normalized confusion matrices for the single-handed and complete sets.}
\label{ConfComplex}
\label{ConfEasy}
\end{figure*}

We present accuracy, precision, recall, and F1-score for within-subject training in Table~\ref{tab1}, Table~\ref{tab2}, and Table~\ref{tab3}.
The confusion matrix for the single-handed gestures (Fig~\ref{EasyWithin}) demonstrates a within-user test accuracy of 98.27\%.
Notably, perfect classification is achieved for seven gestures, while the remaining classes accuracies exceed~90\%.
The confusion matrix for the complete set of gestures (Fig~\ref{ComplexWithin}) yields an overall within-user test accuracy of 98.13\%.
Out of 21 gestures, 16 were classified with 100\% accuracy, and rest with accuracies greater than 90\%.

The performance is evaluated by the Random Forest Classifier models from~\cite{zhang2022real}.
Particularly, we employed a classifier utilizing exclusively statistical features derived from the phase signal (RFC with SP), both statistical features and wavelet coefficients obtained from the phase (RFC with SWP), as well as statistical featues from both phase and RSS (RFC with SPR).
In addition, the model is evaluated using Early Fusion~\cite{calatrava2023light}, Late Fusion~\cite{golipoor2024rfid}, EUIGR~\cite{yu2019rfid}, and GRfid~\cite{zou2016grfid}.
All models were implemented and evaluated across three distinct datasets, constructed as follows:
\\[.2cm]
\noindent\begin{tabular}{lp{6.5cm}}
Dataset \rom{1}: & Gesture data were collected at a distance of 3 meters in Environment A. \\
Dataset \rom{2}: & Gesture data were collected at a distance of 1.5 meters in Environment A.\\
Dataset \rom{3}: & Gesture data were collected at a distance of 1.5 meters in Environment B.
\end{tabular}

Our model consistently outperforms the other models across all datasets (cf. Table~\ref{tab1}, Table~\ref{tab2}, and Table~\ref{tab3}).
Among the random forest models, RFC with SPR, including more features and additional modalities improves the recognition accuracy.
Late Fusion outperforms Early Fusion while EUIGR and GRfid exhibit notably lower performance, with GRfid in particular performing poorly across all metrics (e.g., 30.34\% accuracy).

The confusion matrices in Fig~\ref{EasyLopo} and Fig~\ref{ComplexLopo} depict the performance of our model following leave-one-person-out cross-validation on Dataset \rom{1} for single-handed and complete set of gestures, achieving test accuracies of 94.44\% and 89.28\%.

\subsection{Parameter tuning}
\begin{figure*}
\centering
\begin{subfigure}{0.32\textwidth}
    \centering
    \includegraphics[width=\linewidth]{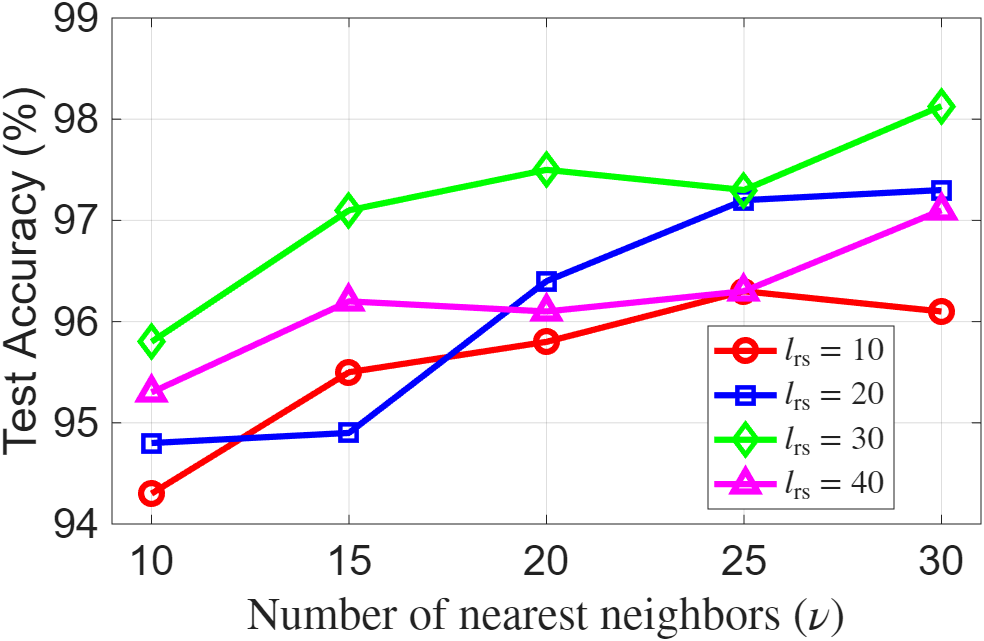}
    \caption{3m distance to the reader}
    \label{3m}
\end{subfigure}
 \hfill
\begin{subfigure}{0.33\textwidth}
    \centering
    \includegraphics[width=\linewidth]{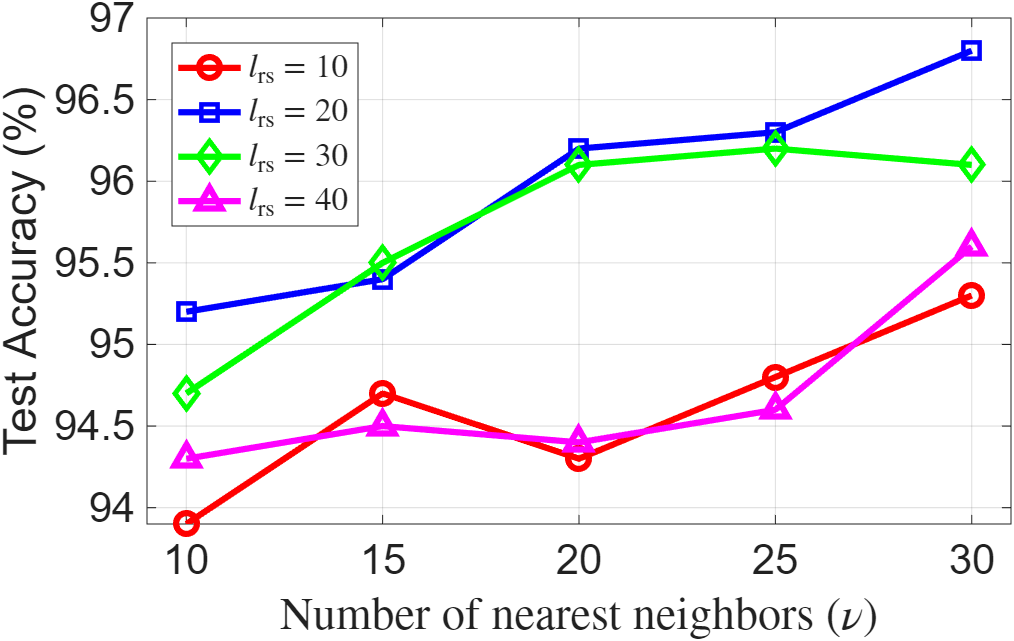}
    \caption{1.5m distance to the reader}
    \label{1.5m}
\end{subfigure}
 \hfill
\begin{subfigure}{0.32\textwidth}
    \centering
    \includegraphics[width=\linewidth]{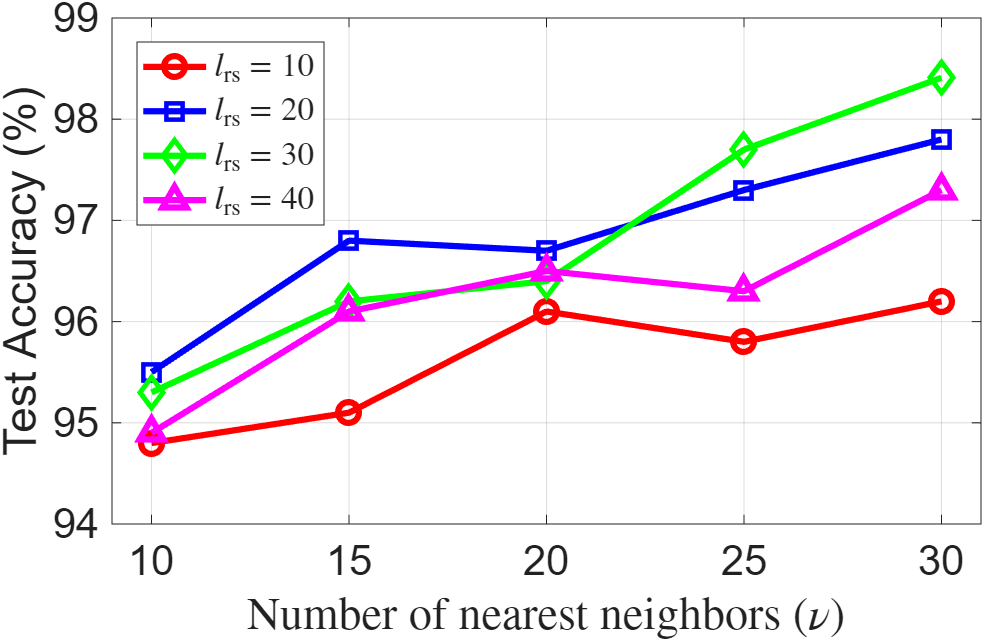}
    \caption{1.5m distance in a different environment}
    \label{1.5m90}
\end{subfigure}
\caption{Test accuracy results for varying combinations of $\nu$ and $l_\text{rs}$}
\label{AccuracyIntermsofk}
\end{figure*}

Fig~\ref{AccuracyIntermsofk} shows the model's performance for different configurations of the top-$\nu$ nearest data points and the resampling length, $l_\text{rs}$.
For datasets \rom{1} and \rom{3}, the highest accuracies were achieved with $l_\text{rs} = 30$ (98.13\% and 98.41\%).
For Dataset \rom{2}, the best performance was obtained with $l_\text{rs} = 20$ (96.82\%). Overall, $l_\text{rs}$ values of 20 and 30 demonstrated superior performance compared to the settings of 10 and 40.

Additionally, increasing $\nu$ generally improved accuracy, with the optimal performance observed at $\nu = 30$.

\subsection{Impact of tag placement and misdetection}
\begin{figure}
\centering
\includegraphics[width=0.8\linewidth]{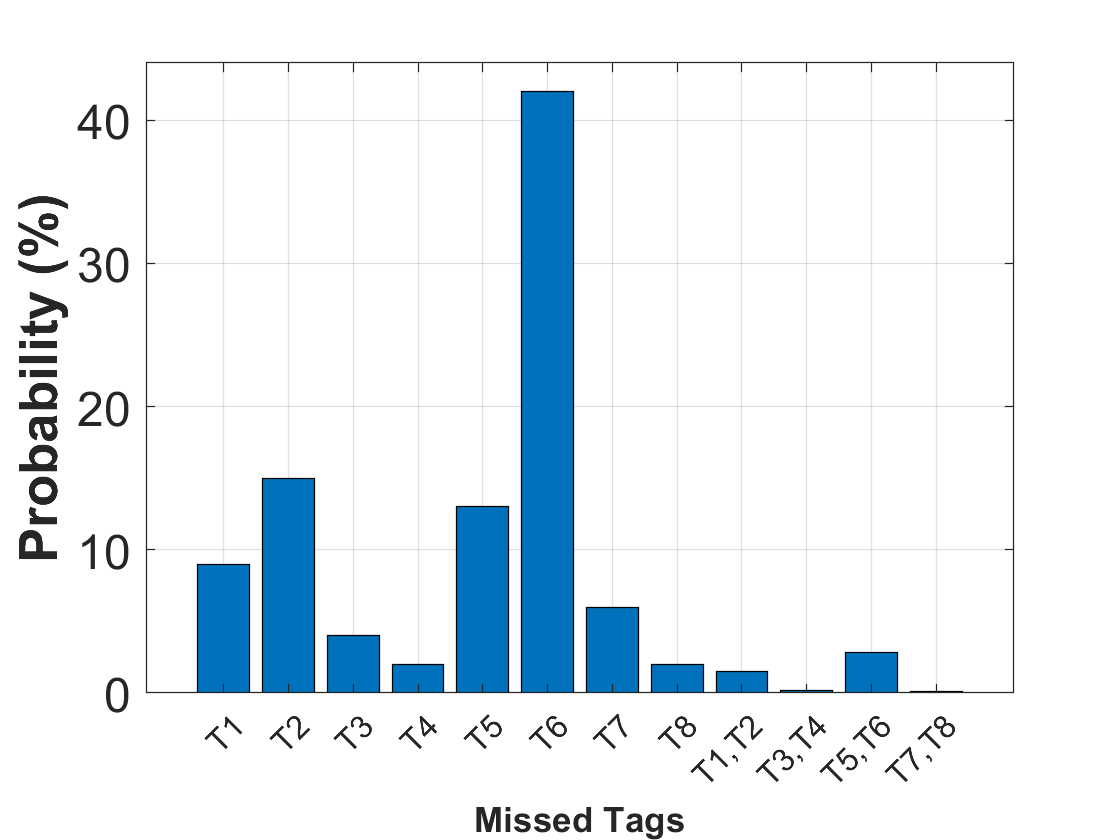}
\caption{Probability of individual tag misdetection and joint tags' misdetection across all gesture executions.}
\label{fig:MisProbablity}
\end{figure}

Fig~\ref{fig:MisProbablity} depicts the proportion of gesture instances in which one or more tags were not detected.
Among all tags, tag~6 exhibited the highest rate of loss (42\%), followed by tag~2 (15\%) due to the placement of these tags at the back of the right and left wrist (cf. Fig~\ref{SetUp}).
Missed detections were also observed for Tags 1 and 5, located on the front side of the wrists, with loss rates of 9\% and 15\%, respectively. In contrast, Tags 3, 4, 7, and 8, which are positioned on the forearms and upper arms, demonstrated significantly lower loss rates.

Notably, as shown in Fig~\ref{fig:MisProbablity}, instances where any pair of tags is lacking data at the same time are rare, so that indeed imputation is well feasible drawing on the samples collected by other tags.

\begin{figure*}
\centering
\begin{subfigure}{0.32\textwidth}
    \centering
    \includegraphics[width=\linewidth]{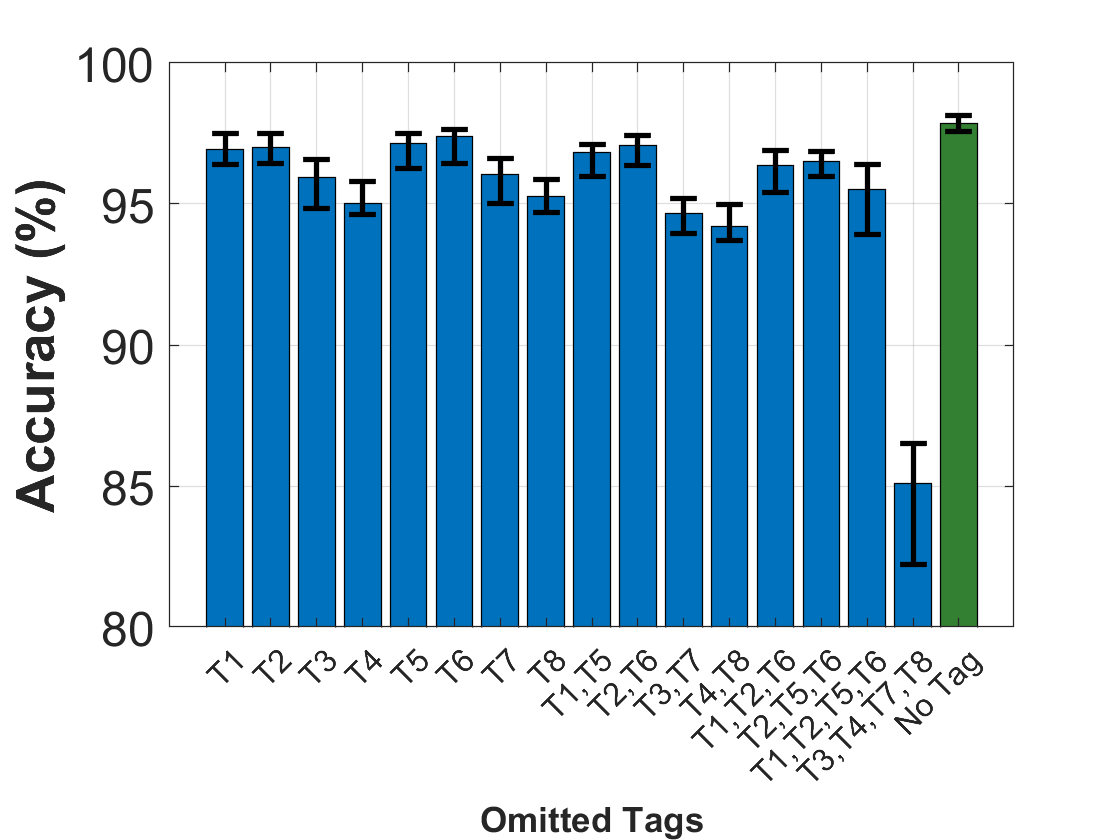}
    \caption{3m distance to the reader}
    \label{3mTagEffect}
\end{subfigure}
 \hfill
\begin{subfigure}{0.32\textwidth}
    \centering
    \includegraphics[width=\linewidth]{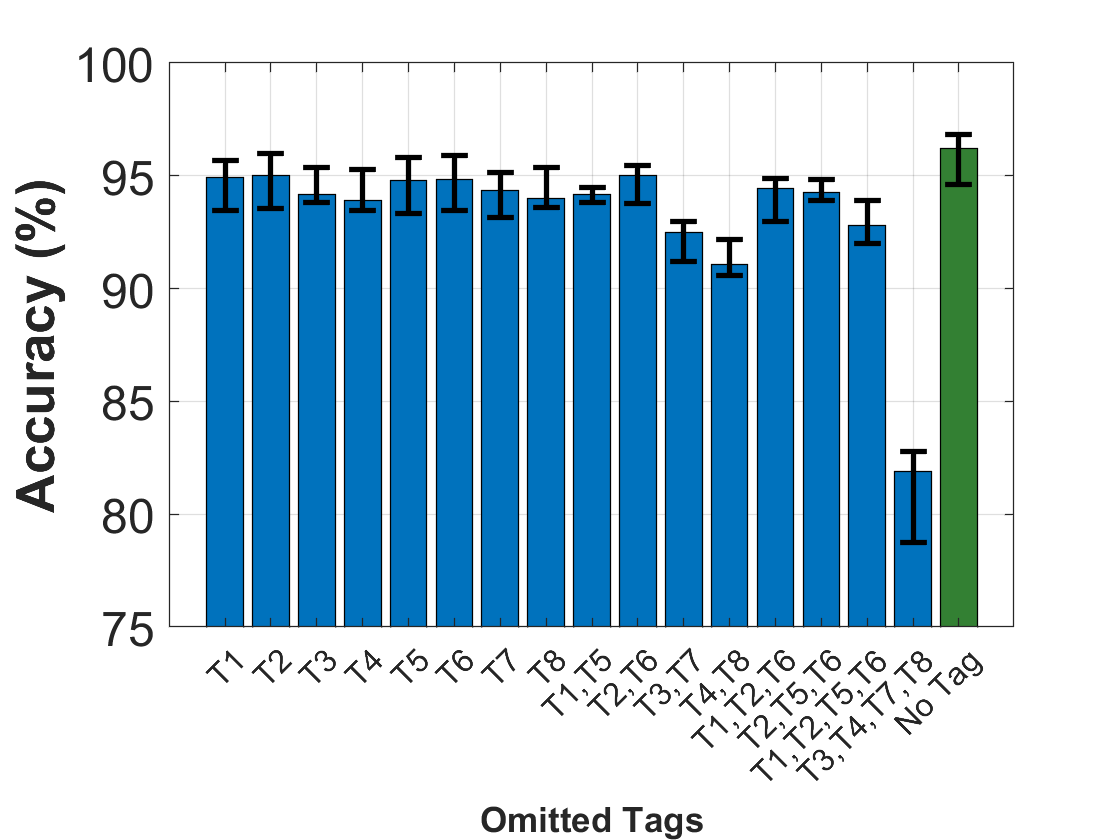}
    \caption{1.5m distance to the reader}
    \label{1.5mTagEffect}
\end{subfigure}
 \hfill
\begin{subfigure}{0.32\textwidth}
    \centering
    \includegraphics[width=\linewidth]{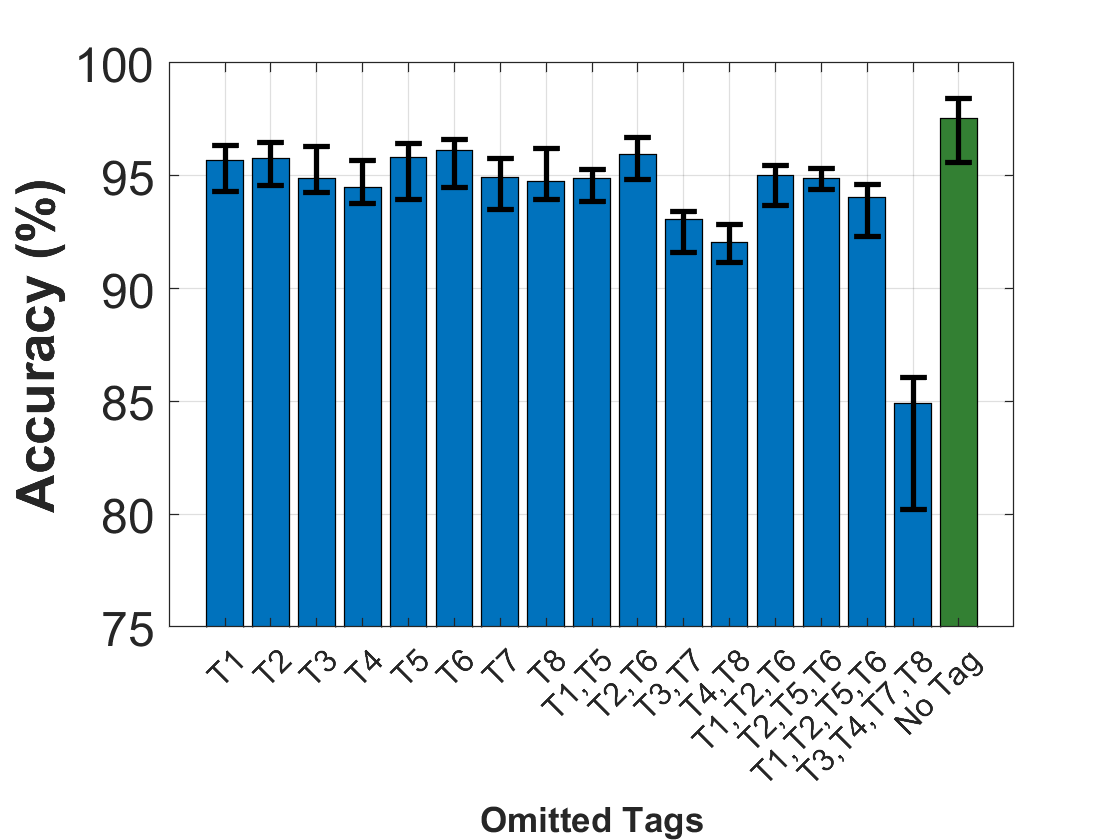}
    \caption{1.5m distance in a different environment}
    \label{1.5m90DegTagEffect}
\end{subfigure}
\caption{Test accuracy under the removal of individual tags and sets of tags. }
\label{AccuracyIntermsofRemovingTags}
\end{figure*}

Fig~\ref{AccuracyIntermsofRemovingTags} evaluates how the system's accuracy changes across all datasets when specific tags or combinations of tags are omitted.
It is evident that certain tags contribute more to the overall performance of the recognition system. Notably, T4 and T8 consistently exhibit the highest individual impact on accuracy. For instance, in Dataset \rom{1} (Fig~\ref{3mTagEffect}), omitting either T4 or T8 leads to a noticeable drop in accuracy from approximately 97.8\% to around 95\%, and further to ~94\% when both are removed. Following T4 and T8, T3 and T7 also demonstrate relatively higher importance compared to tags placed on the wrist, with their combined removal resulting in an accuracy of approximately 94.6\%.
Overall, in all datasets, the "No Tag" condition, where none of the tags are removed, yields the highest accuracy. This indicates that utilizing all eight tags attached to various positions on the garment is beneficial for the recognition system. However, as illustrated in Fig~\ref{fig:MisProbablity}, tag misdetections may occur, leading to intermittent missing values or complete detection failures at certain timestamps. These findings show the importance of our model in robustly handling missing tag data.

\begin{table*}
\caption{Accuracy (Acc.), Precision (Pre.), Recall (Rec.), and F1-score (F1) for all methods}
\begin{subtable}{0.32\textwidth}
\caption{Distance $3~\text{m}$; environment A}
\centering
\setlength\tabcolsep{2.1pt} 
\begin{tabular}{|c|c|c|c|c|}
\hline
\multicolumn{1}{|c|}{\textbf{Method}} &  \textbf{Acc.} & \textbf{Pre.} & \textbf{Rec.} & \textbf{F1} \\ \hline
RFC with SP & $83.56$ & $83.72$ & $83.56$ & $83.42$ \\ \hline
RFC with SWP & $86.26$ & $86.48$ & $86.2$ & $86.07$ \\ \hline
RFC with SPR & $95.25$ & $95.35$ & $95.25$ & $95.23$ \\ \hline
Early Fusion  & $83.62$ & $84.25$ & $83.57$ & $83.34$ \\ \hline
Late Fusion  & $87.13$ & $88.32$ & $87.08$ & $86.91$ \\ \hline
EUIGR & $80.4$ & $78.94$ & $80.07$ & $78.73$ \\ \hline
GRfid & $30.34$ & $31.4$ & $30.16$ & $30.22$ \\ \hline
Our model & $\mathbf{98.13}$ & $\mathbf{98.19}$ & $\mathbf{98.13}$ & $\mathbf{98.13}$ \\ \hline
\end{tabular}
\label{tab1}
\end{subtable}
\hfill
\begin{subtable}{0.32\textwidth}
\caption{Distance $1.5~\text{m}$; environment A}
\centering
\setlength\tabcolsep{2.1pt} 
\begin{tabular}{|c|c|c|c|c|}
\hline
\multicolumn{1}{|c|}{\textbf{Method}} &  \textbf{Acc.} & \textbf{Pre.} & \textbf{Rec.} & \textbf{F1} \\ \hline
RFC with SP & $85.07$ & $86.29$ & $85.02$ & $85.13$ \\ \hline
RFC with SWP & $85.71$ & $86.72$ & $85.71$ & $85.72$ \\ \hline
RFC with SPR & $93.52$ & $94.35$ & $93.52$ & $93.59$ \\ \hline
Early Fusion  & $81.01$ & $86.08$ & $81$ & $81.18$ \\ \hline
Late Fusion  & $89.41$ & $90.18$ & $89.39$ & $89.39$ \\ \hline
EUIGR & $80.37$ & $74.75$ & $80.33$ & $75.97$ \\ \hline
GRfid & $29.39$ & $29.14$ & $29.34$ & $28.89$ \\ \hline
Our model & $\mathbf{96.82}$ & $\mathbf{97.88}$ & $\mathbf{96.8}$ & $\mathbf{97.02}$ \\ \hline
\end{tabular}
\label{tab2}
\end{subtable}
\hfill
\begin{subtable}{0.32\textwidth}
\caption{Distance $1.5~\text{m}$; environment B}
\centering
\setlength\tabcolsep{2.1pt} 
\begin{tabular}{|c|c|c|c|c|}
\hline
\multicolumn{1}{|c|}{\textbf{Method}} &  \textbf{Acc.} & \textbf{Pre.} & \textbf{Rec.} & \textbf{F1} \\ \hline
RFC with SP & $84.12$ & $84.67$ & $84.07$ & $83.96$ \\ \hline
RFC with SWP & $81.26$ & $82.39$ & $81.26$ & $80.93$ \\ \hline
RFC with SPR & $93.80$ & $94.08$ & $93.78$ & $93.71$ \\ \hline
Early Fusion  & $91.13$ & $94.16$ & $91.13$ & $91.56$ \\ \hline
Late Fusion  & $90.15$ & $91.13$ & $90.15$ & $90.15$ \\ \hline
EUIGR & $87.30$ & $83.75$ & $87.27$ & $84.55$ \\ \hline
GRfid & $34.43$ & $36.3$ & $33.87$ & $34.33$ \\ \hline
Our model & $\mathbf{98.41}$ & $\mathbf{98.67}$ & $\mathbf{98.41}$ & $\mathbf{98.39}$ \\ \hline
\end{tabular}
\label{tab3}
\end{subtable}
\end{table*}

\section{Conclusion}
We have explored the use of body-mounted UHF RFID tags for human gesture recognition.
Leveraging a commercial RFID reader and multiple passive tags positioned on the body, we recognized 21 distinct gestures.
We proposed a processing framework to handle partially missing dataframes using linear and exponential interpolation/extrapolation for phase and RSS values.
Fully missing frames were addressed through imputation and proximity-based inference.
The processed data were structured as graph, with each EPC treated as a node and edges formed based on temporal correlations between RSS and phase values across successive timestamps.
This graph representation was then utilized by a self-attention-based graph convolutional network.
We achieved a recognition accuracy of 98.13\%.


%



\ifCLASSOPTIONcaptionsoff
  \newpage
\fi

\bibliographystyle{ieeetr}
\bibliography{ref}

@article{regani2022gwrite,
  title={GWrite: Enabling through-the-wall gesture writing recognition using WiFi},
  author={Regani, Sai Deepika and Wang, Beibei and Hu, Yuqian and Liu, KJ Ray},
  journal={IEEE Internet of Things Journal},
  volume={10},
  number={7},
  pages={5977--5991},
  year={2022},
  publisher={IEEE}
}

@article{yao2023human,
  title={Human Gesture Recognition Based on CT-A Hybrid Deep Learning Model in WiFi Environment},
  author={Yao, Yancheng and Zhao, Chuanxin and Pan, Yahui and Sha, Chao and Rao, Yuan and Wang, Taochun},
  journal={IEEE Sensors Journal},
  year={2023},
  publisher={IEEE}
}

@article{chen2024wignn,
  title={WiGNN: WiFi-Based Cross-Domain Gesture Recognition Inspired by Dynamic Topology Structure},
  author={Chen, Yinan and Huang, Xiaoxia},
  journal={IEEE Wireless Communications},
  year={2024},
  publisher={IEEE}
}

@article{qin2024direction,
  title={Direction-agnostic gesture recognition system using commercial WiFi devices},
  author={Qin, Yuxi and Sigg, Stephan and Pan, Su and Li, Zibo},
  journal={Computer Communications},
  volume={216},
  pages={34--44},
  year={2024},
  publisher={Elsevier}
}

@article{gu2023attention,
  title={Attention-based gesture recognition using commodity wifi devices},
  author={Gu, Yu and Yan, Huan and Zhang, Xiang and Wang, Yantong and Huang, Jinyang and Ji, Yusheng and Ren, Fuji},
  journal={IEEE Sensors Journal},
  volume={23},
  number={9},
  pages={9685--9696},
  year={2023},
  publisher={IEEE}
}

@article{gu2022wigrunt,
  title={WiGRUNT: WiFi-enabled gesture recognition using dual-attention network},
  author={Gu, Yu and Zhang, Xiang and Wang, Yantong and Wang, Meng and Yan, Huan and Ji, Yusheng and Liu, Zhi and Li, Jianhua and Dong, Mianxiong},
  journal={IEEE transactions on human-machine systems},
  volume={52},
  number={4},
  pages={736--746},
  year={2022},
  publisher={IEEE}
}

@article{yu2024mmwave,
  title={A mmWave MIMO Radar-based Gesture Recognition Using Fusion of Range, Velocity, and Angular Information},
  author={Yu, Jih-Tsun and Tseng, Yen-Hsiang and Tseng, Po-Hsuan},
  journal={IEEE Sensors Journal},
  year={2024},
  publisher={IEEE}
}

@article{wu2024lightweight,
  title={A lightweight network with multi-feature fusion for mmWave radar based hand gesture recognition},
  author={Wu, Yajie and Wang, Xiang and Guo, Shisheng and Zhang, Bo and Cui, Guolong},
  journal={IEEE Sensors Journal},
  year={2024},
  publisher={IEEE}
}

@article{jin2023interference,
  title={Interference-robust millimeter-wave radar-based dynamic hand gesture recognition using 2D CNN-transformer networks},
  author={Jin, Biao and Ma, Xiao and Zhang, Zhenkai and Lian, Zhuxian and Wang, Biao},
  journal={IEEE Internet of Things Journal},
  year={2023},
  publisher={IEEE}
}

@article{jin2024rodar,
  title={Rodar: Robust Gesture Recognition Based on mmWave Radar Under Human Activity Interference},
  author={Jin, Can and Meng, Xiangzhu and Li, Xuanheng and Wang, Jie and Pan, Miao and Fang, Yuguang},
  journal={IEEE Transactions on Mobile Computing},
  year={2024},
  publisher={IEEE}
}

@article{qiao2024simple,
  title={Simple and Efficient Gesture Recognition Based on Frequency Modulated Continuous Wave Radar},
  author={Qiao, Lihong and Wang, Zehua and Shu, Yucheng and Xiao, Bin and Luan, Xiao and Shi, Yuhang and Li, Weisheng and Gao, Xinbo},
  journal={IEEE Transactions on Instrumentation and Measurement},
  year={2024},
  publisher={IEEE}
}

@article{salami2022tesla,
  title={Tesla-rapture: A lightweight gesture recognition system from mmwave radar sparse point clouds},
  author={Salami, Dariush and Hasibi, Ramin and Palipana, Sameera and Popovski, Petar and Michoel, Tom and Sigg, Stephan},
  journal={IEEE Transactions on Mobile Computing},
  volume={22},
  number={8},
  pages={4946--4960},
  year={2022},
  publisher={IEEE}
}

@article{nong2024intelligent,
  title={Intelligent sensing technologies based on flexible wearable sensors: A Review},
  author={Nong, Hengchang and Jin, Mingzhe and Pan, Chengcheng and Zhou, Hongyu and Zhang, Chao and Pan, Xuemei and Chen, Yuren and Wei, Xiangning and Lu, Yeping and Zhao, Kaixiao and others},
  journal={IEEE Sensors Journal},
  year={2024},
  publisher={IEEE}
}

@article{tang2024convolutional,
  title={A Convolutional-Transformer based Approach for Dynamic Gesture Recognition of Data Gloves},
  author={Tang, Yingzhe and Pan, Mingzhang and Li, Hongqi and Cao, Xinxin},
  journal={IEEE Transactions on Instrumentation and Measurement},
  year={2024},
  publisher={IEEE}
}

@article{jeon2024applying,
  title={Applying multistep classification techniques with pre-classification to recognize static and dynamic hand gestures using a soft sensor-embedded glove},
  author={Jeon, Sujin and Park, Soyeon and Bae, Joonbum and Lim, Sunghoon},
  journal={IEEE Sensors Journal},
  year={2024},
  publisher={IEEE}
}

@article{wang2024generalizations,
  title={Generalizations of wearable device placements and sentences in sign language recognition with Transformer-based model},
  author={Wang, Qingshan and Zheng, Zhiwen and Wang, Qi and Deng, Dazhu and Zhang, Jiangtao},
  journal={IEEE Transactions on Mobile Computing},
  year={2024},
  publisher={IEEE}
}

@article{calatrava2023light,
  title={Light residual network for human activity recognition using wearable sensor data},
  author={Calatrava-Nicol{\'a}s, Francisco M and Mozos, Oscar Martinez},
  journal={IEEE Sensors Letters},
  year={2023},
  publisher={IEEE}
}

@article{eddy2024discrete,
  title={Discrete Gesture Recognition Using Multi-Modal PPG, IMU, and Single-Channel EMG Recorded at the Wrist},
  author={Eddy, Ethan and Campbell, Evan and C{\^o}t{\'e}-Allard, Ulysse and Bateman, Scott and Scheme, Erik},
  journal={IEEE Sensors Letters},
  year={2024},
  publisher={IEEE}
}

@inproceedings{mubibya2022improving,
  title={Improving human activity recognition using ml and wearable sensors},
  author={Mubibya, Gael S and Almhana, Jalal},
  booktitle={ICC 2022-IEEE International Conference on Communications},
  pages={165--170},
  year={2022},
  organization={IEEE}
}

@article{azarfar2024real,
  title={Real-Time Hand Motion-Modulated Chipless RFID With Gesture Recognition Capability},
  author={Azarfar, Ashkan and Barbot, Nicolas and Perret, Etienne},
  journal={IEEE Transactions on Microwave Theory and Techniques},
  year={2024},
  publisher={IEEE}
}

@inproceedings{azarfar2024hand,
  title={Hand motion-modulated chipless RFID for gesture recognition},
  author={Azarfar, Ashkan and Barbot, Nicolas and Perret, Etienne},
  booktitle={2024 IEEE/MTT-S International Microwave Symposium-IMS 2024},
  pages={349--352},
  year={2024},
  organization={IEEE}
}

@inproceedings{saraf2023survey,
  title={A survey of datasets, applications, and models for IMU sensor signals},
  author={Saraf, Aparajita and Moon, Seungwhan and Madotto, Andrea},
  booktitle={2023 IEEE International Conference on Acoustics, Speech, and Signal Processing Workshops (ICASSPW)},
  pages={1--5},
  year={2023},
  organization={IEEE}
}

@article{li2023semg,
  title={sEMG and IMU Data-based Hand Gesture Recognition Method using Multi-stream CNN with a Fine-tuning Transfer Framework},
  author={Li, Guiyin and Wan, Bo and Su, Kejia and Huo, Jiwang and Jiang, Changhua and Wang, Fei},
  journal={IEEE Sensors Journal},
  year={2023},
  publisher={IEEE}
}

@article{yang2024intelligent,
  title={Intelligent wearable systems: Opportunities and challenges in health and sports},
  author={Yang, Luyao and Amin, Osama and Shihada, Basem},
  journal={ACM Computing Surveys},
  volume={56},
  number={7},
  pages={1--42},
  year={2024},
  publisher={ACM New York, NY}
}

@inproceedings{li2015idsense,
  title={IDSense: A human object interaction detection system based on passive UHF RFID},
  author={Li, Hanchuan and Ye, Can and Sample, Alanson P},
  booktitle={Proceedings of the 33rd Annual ACM Conference on Human Factors in Computing Systems},
  pages={2555--2564},
  year={2015}
}

@article{zhou2017design,
  title={Design and implementation of an RFID-based customer shopping behavior mining system},
  author={Zhou, Zimu and Shangguan, Longfei and Zheng, Xiaolong and Yang, Lei and Liu, Yunhao},
  journal={IEEE/ACM transactions on networking},
  volume={25},
  number={4},
  pages={2405--2418},
  year={2017},
  publisher={IEEE}
}

@inproceedings{shangguan2017enabling,
  title={Enabling gesture-based interactions with objects},
  author={Shangguan, Longfei and Zhou, Zimu and Jamieson, Kyle},
  booktitle={Proceedings of the 15th Annual International Conference on Mobile Systems, Applications, and Services},
  pages={239--251},
  year={2017}
}

@inproceedings{bu2018rf,
  title={RF-Dial: An RFID-based 2D human-computer interaction via tag array},
  author={Bu, Yanling and Xie, Lei and Gong, Yinyin and Wang, Chuyu and Yang, Lei and Liu, Jia and Lu, Sanglu},
  booktitle={IEEE INFOCOM 2018-IEEE conference on computer communications},
  pages={837--845},
  year={2018},
  organization={IEEE}
}

@article{zhang2023rf,
  title={RF-Sign: Position-Independent Sign Language Recognition Using Passive RFID Tags},
  author={Zhang, Haotian and Wang, Lukun and Pei, Jiaming and Lyu, Feng and Li, Minglu and Liu, Chao},
  journal={IEEE Internet of Things Journal},
  year={2023},
  publisher={IEEE}
}

@article{zhang2024sign,
  title={Sign Language Recognition Based on CNN-BiLSTM Using RF Signals},
  author={Zhang, Yajun and Wang, Yuankang and Li, Feng and Yu, Weiqian and Wang, Congcong and Jiang, Ying},
  journal={IEEE Access},
  year={2024},
  publisher={IEEE}
}

@article{xu2023rf,
  title={RF-CSign: A Chinese Sign Language Recognition System Based on Large Kernel Convolution and Normalization-Based Attention},
  author={Xu, Huanyuan and Zhang, Yajun and Yang, Zhixiong and Yan, Haoqiang and Wang, Xingqiang},
  journal={IEEE Access},
  volume={11},
  pages={133767--133780},
  year={2023},
  publisher={IEEE}
}

@article{dian2020towards,
  title={Towards domain-independent complex and fine-grained gesture recognition with RFID},
  author={Dian, Cao and Wang, Dong and Zhang, Qian and Zhao, Run and Yu, Yinggang},
  journal={Proceedings of the ACM on Human-Computer Interaction},
  volume={4},
  number={ISS},
  pages={1--22},
  year={2020},
  publisher={ACM New York, NY, USA}
}

@article{merenda2022edge,
  title={Edge machine learning techniques applied to RFID for device-free hand gesture recognition},
  author={Merenda, Massimo and Cimino, Giuseppe and Carotenuto, Riccardo and Della Corte, Francesco G and Iero, Demetrio},
  journal={IEEE Journal of Radio Frequency Identification},
  volume={6},
  pages={564--572},
  year={2022},
  publisher={IEEE}
}

@article{zhang2022real,
  title={Real-time and accurate gesture recognition with commercial RFID devices},
  author={Zhang, Shigeng and Ma, Zijing and Yang, Chengwei and Kui, Xiaoyan and Liu, Xuan and Wang, Weiping and Wang, Jianxin and Guo, Song},
  journal={IEEE Transactions on Mobile Computing},
  volume={22},
  number={12},
  pages={7327--7342},
  year={2022},
  publisher={IEEE}
}

@inproceedings{wang2018multi,
  title={Multi-touch in the air: Device-free finger tracking and gesture recognition via COTS RFID},
  author={Wang, Chuyu and Liu, Jian and Chen, Yingying and Liu, Hongbo and Xie, Lei and Wang, Wei and He, Bingbing and Lu, Sanglu},
  booktitle={IEEE INFOCOM 2018-IEEE conference on computer communications},
  pages={1691--1699},
  year={2018},
  organization={IEEE}
}

@article{zou2016grfid,
  title={GRfid: A device-free RFID-based gesture recognition system},
  author={Zou, Yongpan and Xiao, Jiang and Han, Jinsong and Wu, Kaishun and Li, Yun and Ni, Lionel M},
  journal={IEEE Transactions on Mobile Computing},
  volume={16},
  number={2},
  pages={381--393},
  year={2016},
  publisher={IEEE}
}

@article{cheng2019air,
  title={In-air gesture interaction: Real time hand posture recognition using passive RFID tags},
  author={Cheng, Kang and Ye, Ning and Malekian, Reza and Wang, Ruchuan},
  journal={IEEE access},
  volume={7},
  pages={94460--94472},
  year={2019},
  publisher={IEEE}
}

@article{xie2017multi,
  title={Multi-touch in the air: Concurrent micromovement recognition using RF signals},
  author={Xie, Lei and Wang, Chuyu and Liu, Alex X and Sun, Jianqiang and Lu, Sanglu},
  journal={IEEE/ACM Transactions on Networking},
  volume={26},
  number={1},
  pages={231--244},
  year={2017},
  publisher={IEEE}
}

@inproceedings{golipoor2024rfid,
  title={RFID-based Human Activity Recognition Using Multimodal Convolutional Neural Networks},
  author={Golipoor, Sahar and Sigg, Stephan},
  booktitle={2024 IEEE 29th International Conference on Emerging Technologies and Factory Automation (ETFA)},
  pages={1--6},
  year={2024},
  organization={IEEE}
}

@article{clarke2006radio,
  title={Radio frequency identification (RFID) performance: the effect of tag orientation and package contents},
  author={Clarke, Robert H and Twede, Diana and Tazelaar, Jeffrey R and Boyer, Kenneth K},
  journal={Packaging Technology and Science: An International Journal},
  volume={19},
  number={1},
  pages={45--54},
  year={2006},
  publisher={Wiley Online Library}
}

@inproceedings{vasisht2018body,
  title={In-body backscatter communication and localization},
  author={Vasisht, Deepak and Zhang, Guo and Abari, Omid and Lu, Hsiao-Ming and Flanz, Jacob and Katabi, Dina},
  booktitle={Proceedings of the 2018 Conference of the ACM Special Interest Group on Data Communication},
  pages={132--146},
  year={2018}
}

@inproceedings{yu2019rfid,
  title={RFID based real-time recognition of ongoing gesture with adversarial learning},
  author={Yu, Yinggang and Wang, Dong and Zhao, Run and Zhang, Qian},
  booktitle={Proceedings of the 17th Conference on Embedded Networked Sensor Systems},
  pages={298--310},
  year={2019}
}

@inproceedings{golipoor2024environment,
  title={Environment and Person-independent Gesture Recognition with Non-static RFID Tags Leveraging Adaptive Signal Segmentation},
  author={Golipoor, Sahar and Sigg, Stephan},
  booktitle={2024 IEEE 29th International Conference on Emerging Technologies and Factory Automation (ETFA)},
  pages={1--8},
  year={2024},
  organization={IEEE}
}

@inproceedings{golipoor2023accurate,
  title={Accurate RF-sensing of complex gestures using RFID with variable phase-profiles},
  author={Golipoor, Sahar and Sigg, Stephan},
  booktitle={2023 IEEE 32nd International Symposium on Industrial Electronics (ISIE)},
  pages={1--4},
  year={2023},
  organization={IEEE}
}

@article{schafer2011savitzky,
  title={What is a savitzky-golay filter?[lecture notes]},
  author={Schafer, Ronald W},
  journal={IEEE Signal processing magazine},
  volume={28},
  number={4},
  pages={111--117},
  year={2011},
  publisher={IEEE}
}

@article{marr1980theory,
  title={Theory of edge detection},
  author={Marr, David and Hildreth, Ellen},
  journal={Proceedings of the Royal Society of London. Series B. Biological Sciences},
  volume={207},
  number={1167},
  pages={187--217},
  year={1980},
  publisher={The Royal Society London}
}

@article{obiri2024survey,
  title={A Survey of LoRaWAN-integrated Wearable Sensor Networks for Human Activity Recognition: Applications, Challenges and Possible Solutions},
  author={Obiri, Nahshon Mokua and Van Laerhoven, Kristof},
  journal={IEEE Open Journal of the Communications Society},
  year={2024},
  publisher={IEEE}
}

@article{yin2024systematic,
  title={A systematic review of human activity recognition based on mobile devices: overview, progress and trends},
  author={Yin, Yafeng and Xie, Lei and Jiang, Zhiwei and Xiao, Fu and Cao, Jiannong and Lu, Sanglu},
  journal={IEEE Communications Surveys \& Tutorials},
  volume={26},
  number={2},
  pages={890--929},
  year={2024},
  publisher={IEEE}
}

@inproceedings{baba2025human,
  title={Human Activity Recognition Scheme Based on Time-Series Analysis of Multi-Antenna AoA Data},
  author={Baba, Kensuke and Naito, Katsuhiro},
  booktitle={2025 IEEE International Conference on Consumer Electronics (ICCE)},
  pages={1--6},
  year={2025},
  organization={IEEE}
}

@article{miao2025wi,
  title={Wi-Fi Sensing Techniques for Human Activity Recognition: Brief Survey, Potential Challenges, and Research Directions},
  author={Miao, Fucheng and Huang, Youxiang and Lu, Zhiyi and Ohtsuki, Tomoaki and Gui, Guan and Sari, Hikmet},
  journal={ACM Computing Surveys},
  volume={57},
  number={5},
  pages={1--30},
  year={2025},
  publisher={ACM New York, NY}
}

@article{zahid2024comprehensive,
  title={A Comprehensive Study of Chipless RFID Sensors for Healthcare Applications},
  author={Zahid, Muhammad Noaman and Gaofeng, Zhu and Sadiq, Touseef and ur Rehman, Hameed and Anwar, Muhammad Shahid},
  journal={IEEE Access},
  year={2024},
  publisher={IEEE}
}

@article{kong2024survey,
  title={A survey of mmwave radar-based sensing in autonomous vehicles, smart homes and industry},
  author={Kong, Hao and Huang, Cheng and Yu, Jiadi and Shen, Xuemin},
  journal={IEEE Communications Surveys \& Tutorials},
  year={2024},
  publisher={IEEE}
}

@article{wen2024survey,
  title={A survey on integrated sensing, communication, and computation},
  author={Wen, Dingzhu and Zhou, Yong and Li, Xiaoyang and Shi, Yuanming and Huang, Kaibin and Letaief, Khaled B},
  journal={IEEE Communications Surveys \& Tutorials},
  year={2024},
  publisher={IEEE}
}
\balance
\end{document}